\documentclass[a4paper,fleqn,usenatbib]{mnras}
\usepackage{epstopdf}
\usepackage{longtable}
\usepackage{graphics,epsfig,psfig,graphicx,multirow}
\usepackage{hyperref}
\usepackage{amsmath}
\usepackage{amssymb}	
\usepackage{pdflscape}
\usepackage{footnote}

\def \be{\begin{equation}}
\def \ee{\end{equation}}
\def \bea{\begin{eqnarray}}
\def \eea{\end{eqnarray}}
\def\etal{{et al.}~}

\def\ltsima{$\; \buildrel < \over \sim \;$}
\def\simlt{\lower.5ex\hbox{\ltsima}}
\def\gtsima{$\; \buildrel > \over \sim \;$}
\def\simgt{\lower.5ex\hbox{\gtsima}}

\usepackage{color}
\usepackage[normalem]{ulem}

\definecolor{darkgreen}{rgb}{0.13, 0.55, 0.13}

\newcommand{\aref}[1]{\hyperref[#1]{Appendix~\ref{#1}}}

\newcommand{\rion}[2]{{\ensuremath{\mbox{\rm #1$\,${\small\uppercase\expandafter{\romannumeral#2\relax}}}}}}

\usepackage{newtxtext,newtxmath}
\usepackage[T1]{fontenc}
\usepackage{ae,aecompl}

\usepackage{color}
\usepackage[normalem]{ulem}
\definecolor{darkgreen}{rgb}{0.13, 0.55, 0.13}

\definecolor{brown}{rgb}{0.59, 0.29, 0.0}

\title[He and N enrichment in WNL stars]{Helium and Nitrogen Enrichment in Massive Main Sequence Stars: Mechanisms and Implications for the Origin of WNL Stars}
\author[A. Roy et al.]{Arpita Roy$^{1,2}$\thanks{E-mail: arpita.roy1016@gmail.com},
Ralph S. Sutherland$^{1,2}$,
Mark R. Krumholz$^{1,2}$,
Alexander Heger$^{2,3,4,5,6}$,
\newauthor
Michael A. Dopita$^{1,2\thanks{deceased}}$
\\
$^{1}$Research School of Astronomy and Astrophysics, Australian National University, Cotter Road, Weston Creek, ACT 2611, Australia.\\
$^{2}$ARC Centre of Excellence for All-sky Astrophysics in 3 Dimensions (ASTRO 3D).\\
$^{3}$School of Physics and Astronomy, Monash Centre for Astrophysics, 19 Rainforest walk, Monash University, VIC 3800, Australia.\\
$^{4}$Tsung-Dao Lee Institute, Shanghai 200240, China.\\
$^{5}$ OzGrav: Australian Research Council Centre of Excellence for
Gravitational Wave Discovery, Clayton, VIC 3800, Australia.\\
$^{6}$Joint Institute for
Nuclear Astrophysics – Center for the Evolution of the Elements.\\
}

\date{Accepted XXX. Received YYY; in original form ZZZ}

\pubyear{2019}

\hypersetup{draft} 
\begin{document}
\label{firstpage}
\pagerange{\pageref{firstpage}--\pageref{lastpage}}
\maketitle

\begin{abstract}
The evolutionary paths taken by massive stars with $M \gtrsim 60 \, \mathrm{M}_\odot$ remain substantially uncertain. They begin their lives as main sequence (MS) O-stars. Depending on their masses, rotation rates, and metallicities, they can then encounter a wide range of evolutionary states with an equally broad set of possible surface compositions and spectral classifications. We present a new grid of calculations for the evolution of such stars that covers a broad range in mass, M/M$_\odot = 60$ to $150$, rotation rate, $v \, / \, v_{\rm crit} = 0$ to $0.6$, metallicity, $[\mathrm{Fe}/\mathrm{H}] = -4$ to $0$, and $\alpha$-element enhancement, $[\alpha/\mathrm{Fe}] = 0$ to $0.4$. 
We show that rotating stars undergo rotationally-induced dredge-up of nucleosynthetic products, mostly He and N, to their surfaces while still on the MS. Non-rotating metal-rich stars also reveal the products of nucleosynthesis on their surfaces because even modest amounts of mass loss expose their ``fossil" convective cores: regions that are no longer convective, but which were part of the convective core at an early stage in the star's evolution. Thus surface enhancement of He and N is expected for rotating stars at all metallicities, and for non-rotating stars if they are relatively metal-rich. We calculate a stellar atmosphere for a representative model from our grid, properly accounting for He- and N-enhancement, and show that the resulting spectrum provides a good match to observed WNL stars, strongly suggesting that the  physical mechanisms we have identified are the ultimate cause of the WNL phase.
\end{abstract}

\begin{keywords}
 stars: abundances-- stars: massive -- stars: mass-loss -- ISM: abundances -- galaxies: high-redshift -- galaxies: ISM
\end{keywords}


\section{Introduction}
\label{intro}
Wolf-Rayet (WR) stars are rare, but represent one of the most important phases of stellar evolution. Spectroscopically, they are distinguished by the presence of strong, broad emission lines, which are associated with winds carrying mass fluxes an order of magnitude larger than those of ordinary O stars. Most WR stars are believed to represent the late evolutionary phases of massive O stars; they have masses $\sim 10- 25 \, \mathrm{M}_\odot$ and based on number counts their average lifetime must occupy only 10\% of the O star's typical $\sim 5$ Myr lifetime \citep{meynet2005, crowther2007}. Among the subtypes of WR stars, the WN group is characterised by strong helium and nitrogen lines. The WN group can be further broken into subclasses such as Early Type WN (WNE) stars (WN2 to WN5) and Late Type WN (WNL) stars (WN7 to WN9) based on the line ratios of N~\textsc{iii}$-$\textsc{iv} and He~\textsc{i}$-$\textsc{ii}. A fraction of WNL stars, known as WNh, show hydrogen emission lines, from which the inferred hydrogen surface mass fraction can be as high as X$_{\rm H}\sim 50$\% \citep{crowther1995, crowther1998}. However, the mapping between surface abundances and spectra is complex, and depends substantially on the properties of the stellar wind. Consequently, it is not entirely clear how to connect these measured surface abundances with evolutionary state, a situation that led \citet{meynet2003} to introduce the eWNL and eWNE classifications, which are based on evolutionary phase rather than spectroscopy. In this scheme the eWNL phase is conventionally taken to correspond to stars with enhanced surface He and depleted, but not entirely absent, H, whereas eWNEs are taken to be stars with whose surfaces are predominantly He by mass. While these definitions are intended to match the spectroscopic ones as closely as possible, their level of fidelity remains significantly uncertain. 

\begin{table*}
\centerline{
\begin{tabular}{cccc}
\hline\hline
WN-subtype & Reference & Proposed definition & Proposed evolutionary state \\
\hline
\multirow{4}{*}{WNL} & \citet{schaerer1992} & X$_{\rm H} \sim 10-50\%$ & Shell H burning \\
& \citet{meynet2005} & X$_{\rm H} \leq 50\%$, Y$_{\rm He}\geq 40\%$ & - \\
& \citet{crowther2007} & X$_{\rm H} \sim 5 - 25\%$ & Core H burning \\ 
& This work & Y$_{\rm He} = 40 - 90\%$ & - \\
\hline
\multirow{3}{*}{WNE} & \citet{schaerer1992} & X$_{\rm H} \leq 10\%$, Y$_{\rm He} \geq 90\%$ &
Core He burning \\
& \citet{crowther2007} & X$_{\rm H} \lesssim 5\%$ & Core He burning\\
& This work & Y$_{\rm He} > 90\%$ & - \\
\hline\hline
\end{tabular}
}
\caption{
\label{tab:def_tab}
Proposed definitions of WN-subtypes in terms of surface abundance, and proposed identifications between subtypes and internal evolutionary phase.
}
\end{table*}

Stars in the WR spectral classes, and the WN sub-type in particular, are commonly observed in young, massive star clusters.  For example, the Arches Cluster \citep{cotera1992, cotera1995, cotera1996, figer1995, nagata1995} contains $\sim 30$ stars classified by \citet{figer2002} as  ``Hydrogen-burning WR stars on the main sequence", which later authors have classified as WN due to their weak or non-existent H features, and their correspondingly high abundances of He and N. The 30 Doradus region of the Large Magellanic Cloud (LMC) contains $\sim 20$ stars classified as WRs \citep{evans2011}, whereas the region around HD 97950 in NGC 3603 contains only a handful \citep{crowther1998}. 

The physical mechanisms by which ordinary massive stars evolve into WR stars is a subject of ongoing study. Classical WR stars are produced by the loss of the hydrogen envelope via strong winds or Roche lobe overflow in close binary systems, leading to exposure of the bare helium core \citep{abbott1987, chiosi1986}. WNL stars, however, do not always originate following a classical WR wind-driven mass-loss process.  Instead, they can be produced by the dredge-up of the CNO-cycle products to the stellar surface by rotational mixing \citep{heger2000, meynet2000, meynet2005} for moderately or fast rotating  stars. 
\citet{yusof2013} modelled a grid of masses ranging from 120$- 500$ 
$\mathrm{M}_\odot$ for solar, LMC (Large Magellanic Cloud) and SMC (Small Magellanic Cloud) metallicities and both for rotating and non-rotating cases in order to investigate the evolution and fates of nearby very massive stars (VMS). They found that, because these stars have very massive convective cores at the main sequence, their evolution is mostly chemically-homogenous regardless of the effects of rotational mixing. Similarly, \citet{kohler2015} modelled a grid of masses ranging from $70 - 500$ M$_\odot$ with rotational velocities $0- 550$ km/s for a chemical composition tailored to match the LMC. They found that rapid rotators are chemically homogeneous, and the very massive stars ($\gtrsim 160$ M$_\odot$) also  become equally homogeneous due to stellar wind mass-loss irrespective of their rotation rates. \citet{limongi2018} modeled a grid of massive stars ($13- 120$ M$_\odot$) for three rotation rates (0, $150$, $300$ km/s) for four metallicities ([Fe/H] $= 0$, $-1$, $-2.0$, and $-3.0$) until the end of presupernova phase. They found that for the non-rotating solar metallicity stars, all massive stars beyond $\sim 17$ M$_\odot$ explode as WR stars, while for the rotating case, all stars explode as WR stars.
\citet{szecsi2015} modelled the main-sequence evolution of 9$- 300 \, \mathrm{M}_\odot$ stars with rotational velocities of $0- 900$ km s$^{-1}$ to study the metallicity- and rotation-dependence of massive stellar evolution at metallicities relevant to the Early Universe. They found that moderately fast rotators have quasi-chemically homogeneous evolution due to effective rotational mixing. This causes the stars to produce He~\textsc{ii} ionizing flues that match observations of low-metallicity He~\textsc{ii} galaxies.  \citet{choi2017} modelled very massive stars for a variety of rotation rates, and found ionizing luminosities consistent with those observed from nuclear star-clusters. They concluded that moderately fast rotators (40\% of the break-up velocity) make a significant contribution to the ionizing photon budget in low-metallicity environments.

A simple picture in which WNL stars are solely the result of rotational mixing, however, encounters the problem that WNL-like abundance patterns have also been observed on the surfaces of slow rotators or non-rotating massive stars. \citet{herrero2000} observed seven Galactic O-stars of type O6 and earlier, three of which show signatures of high helium abundance ($\epsilon = N_{\rm He}/(N_{\rm H}+N_{\rm He}) \approx 0.14$, corresponding to a helium mass fraction of $\sim 35- 40 \%$) but also very slow rotation rates of $\sim 100- 120$ km s$^{-1}$. They estimated initial masses of $\sim 114$, 142, and 97 $\mathrm{M}_\odot$ respectively for these stars. Rotational mixing cannot explain the high helium abundance of such a slowly rotating star, leading \citet{meynet2000} to hypothesise that these slow rotators might have originated from the fast rotators that dredged He and N to their surfaces while on the main sequence, and subsequently lost angular momentum via their winds after leaving the main sequence. More recently, \citet{vink2017} found that 4 in a sample of 39 WR stars in the LMC have surfaces enriched with heavy metals despite of being slow-rotators. They also concluded that this observation challenges  rotationally-induced chemically homogenous evolution (CHE) models, and instead hints at a different dredge-up mechanism specific to very massive stars (VMS).

Not surprisingly, given the uncertainty about their origins, there is significant disagreement in the literature about the relationship between WN-subtypes and stars' initial masses and internal evolutionary states. For example, \citet{meynet1994} proposes that WNL stars have initial masses of $65- 110$ $\mathrm{M}_\odot$, and \citet{langer1994}, \citet{crowther1995}, and
\citet{crowther2007} hypothesise that more than $50$ \% of the more massive ($\geqslant 75 \, \mathrm{M}_\odot$ at solar metallicity) WNL stars are still core hydrogen burning, whereas \citet{schaerer1992} argues that WNL-like surface mass fractions arise during the evolutionary phase when stars are H shell burning. \citet{crowther2007} further suggests that only a small percentage $\sim 15\pm 10$\% of less-massive ($\leqslant 25 \, \mathrm{M}_\odot$) stars at solar metallicities show WNL-like features during the core-He burning phase. Both \citet{schaerer1992} and \citet{crowther2007} propose that stars with WNE-like surface mass-fractions are in the core He-burning phase. \citet{meynet2005} do not suggest any clean mapping between WN subtype and internal evolutionary state. We summarise some of the proposed evolutionary sequences in \autoref{tab:def_tab} for massive stars of $60$ to $150 \, \mathrm{M}_\odot$.

In this paper, we show that the exposure of ``fossil" convective cores in non-rotating stars or slow rotators can significantly enhance the surface mass-fraction of CNO-cycle byproducts to a magnitude almost equivalent to the surface enrichments caused by the  rotational dredge-up in moderately/fast rotators. Therefore, high surface mass-fraction of helium and nitrogen in the most massive slow-rotators at high metallicity, [Fe/H]$\ge 0.0$, is an inevitable phenomenon. As a result, WNL stars can appear at a wide range of rotation rates and metallicities. This paper is organised as follows:  in \autoref{methods_sec}, we describe the numerical methods adopted in our stellar evolution models; we report the results in \autoref{results_sec}, and we discuss their implications and caveats in \autoref{discussions_sec}. We conclude and summarize our results in \autoref{conclusions_sec}.


\section{Method: Stellar evolution calculation}
\label{methods_sec}
We use MESA \citep{paxton2011, paxton2013, paxton2015} version 9793 for all our calculations of stellar evolution. Except as described below, we configure MESA to include all the same physical processes and parameters values as used in the MESA Isochrone Stellar Tracks - $\rm I$(MIST-$\rm I$) track library described by \citet{choi2016}. The only differences between our models and the MIST-I setup is in our treatment of mixing mechanisms and  our inclusion of non-solar-scaled abundances. We discuss these differences in \autoref{mixing_subsec} and \autoref{grid_subsec}.

\subsection{Mixing mechanisms}
\label{mixing_subsec}
Mixing in MESA is described in terms of two diffusion coefficients, one of which, $D$, describes diffusion of composition and the other of which, $\gamma$, describes diffusion of angular momentum via effective (turbulent) viscosity. These need not be the same. The diffusion coefficients are approximated as the sum of a series of terms, each describing the diffusion rate associated with some instability or process. Some of these diffusion coefficients are related to rotation or magnetic fields (which also require rotation to transport anything), whereas others are not. The non-rotational processes are convection (conv), convective overshoot (ovr), thermohaline (thrm), and semiconvective (sem) mixing. The rotational / magnetic processes are dynamical shear instability (DSI), Solberg-Hoiland instability (SHI), secular shear instability (SSI), Eddington-Sweet (ES) circulation, Goldreich-Schubert-Fricke (GSF) instability, and the instability caused by magnetic torques by dynamo-generated fields referred to as Spruit-Taylor (ST) dynamo.

We calculate $D$ and $\gamma$ following \citet{choi2016}, with the exception that we also include ST folllowing \citet{heger2000, heger2005}. Specifically, we calculate $D$ and $\gamma$ as: 
\begin{eqnarray}
D &=& D_{\rm {conv}}+D_{\rm {ovr}}+D_{\rm {thrm}}+D_{\rm {sem}}+f_c \nonumber\\
&& \times (D_{\rm {DSI}}+D_{\rm {SHI}}+D_{\rm {SSI}}+D_{\rm {ES}}+D_{\rm {GSF}} \nonumber\\
&& +D_{\rm {ST}})
\label{eq:D_diff}
\\
\gamma &=& D_{\rm {conv}}+D_{\rm {ovr}}+D_{\rm {thrm}}+D_{\rm {sem}}\nonumber \\
&& +D_{\rm {DSI}}+D_{\rm {SHI}}+D_{\rm {SSI}}+D_{\rm {ES}}+D_{\rm {GSF}} 
\nonumber\\
&& +D_{\rm {ST}}\, ,
\label{eq:nu}
\end{eqnarray}
where $f_{\rm c}$ is a parameter between 0 and 1 the describing the ratio of the diffusion coefficient to the turbulent viscosity. We adopt as a fiducial value $f_{\rm c}=1/30$, following \citet{choi2016} (also see references therein).
We calculate $D_{\rm {ST}}$ following equation 41 of \citet{spruit2002}, which follows the standard approach used for high-mass stellar evolution in MESA \citep{paxton2013},  KEPLER \citep{heger2005} and STERN \citep{petrovic2005}. For the details of mixing mechanisms and angular momentum transport in MIST, we refer readers to \citet{choi2016}.

Because the values of $D$ and $\gamma$ depend on parameterised models for various processes, the exact transport rates are quite uncertain. We explore the sensitivity of our final results to this uncertainty in \aref{conv_rot} and \aref{conv_ovr}, and we discuss the implications of these experiments in \autoref{param_sensitivity_sec}.

\subsection{Treatment of stellar winds and mass loss}
\label{ssec:winds}

Our treatment of mass loss via stellar winds match that of \citet{choi2016}, which is based on the ``Dutch'' mass loss recipe that is standard in MESA \citep{paxton2013}. Specifically, for O stars we adopt the \citet{vink2001} model for metallicity-dependent winds, and for WR stars we use the \citet{nugis2000} model in which the mass loss rate depends on both luminosity and surface composition. Mass loss is enhanced by rotation following the prescription described in \citet{choi2016}. Mass loss is, as usual, a dominant uncertainty in models for the evolution of massive stars \citep{smith2014}, and we explore the sensitivity of our results to the adopted prescriptions for wind mass loss in \aref{mass_loss_subsec}, and discuss the results of these experiments in \autoref{param_sensitivity_sec}. We note that a number of recent observational studies \citep{grafener2011, 
vink2012, besten2014} suggest that stars that lie close to the Eddington limit have stronger winds than those predicted by the \citet{vink2001} prescription, and that these results have been incorporated into some more recent empirical mass-loss prescriptions \citep[e.g.,][]{hainich2015}. Given the large uncertainties, a detailed comparison of alternative mass-loss prescriptions is beyond the scope of this paper; we instead prefer to characterise our uncertainties based on mass loss using the simpler experiments presented in \aref{mass_loss_subsec}, where we simply parameterise the factor by which the mass loss rate could be larger or smaller than our fiducial choice.

\subsection{Model grid}
\label{grid_subsec}

We calculate models for a three-dimensional grid in initial mass, abundance, and rotation rate as follows:
\begin{itemize}
\item \textit{Initial mass}: We consider initial masses of 60$- 150$ $\mathrm{M}_\odot$ with a spacing $\Delta M = 10$ $\mathrm{M}_\odot$. We choose these masses to provide dense coverage of the range of masses suspected to be the progenitor population of WNL stars.

\item {\it Abundance}: We use [Fe/H] values of 0.0, $-1.0$, $-2.0$, $-4.0$ with respect to the protosolar metallicity \citep{asplund2009}. Unlike MIST-I (though see MIST-II, \citet{dotter}), we do not assume that all elements scale with Fe. Instead, [$\alpha$/Fe] has been observed to vary systematically with [Fe/H] (see Figure 2 of \citealt{nicholls2017}). We adopt the scaling given in that paper, namely that below [Fe/H] $= -1.0$, [$\alpha$/Fe] is constant at $+0.4$, and in the range $-1.0 < [\mathrm{Fe}/\mathrm{H}] < 0.0$ it decreases linearly to  $[\alpha/\mathrm{Fe}] = 0.0$ at solar abundance, and continues decreasing with the same slope at super-solar abundances. In our calculations we implement non-solar $[\alpha/\mathrm{Fe}]$ values by varying the initial abundances of all $\alpha$-elements (O, Ne, Mg, Si, S, Ca, and Ti) by the same factor. We calculate the initial helium abundance adopting a scaling of $\Delta Y/ \Delta Z=1.5$ with the primordial helium abundance $Y_{\rm p}=0.249$ \citep{planck2016} and metallicity Z, and calculate $Y$ as,
\begin{equation}
 Y=Y_{\rm{p}}+\left({{Y_{\odot \rm{, protosolar}}-Y_{\rm {p}}} \over {Z_{\odot \rm{, protosolar}}}}\right) Z \, ,
\label{eq:Y} 
\end{equation}
with $Y_{\odot \rm{, protosolar}}=0.2703$ and $Z_{\odot \rm{, protosolar}}=0.0142$ \citep{asplund2009}, where
\begin{equation} 
{{Y_{\odot \rm{, protosolar}}-Y_{\rm {p}}} \over {Z_{\odot \rm{, protosolar}}}}=\Delta Y/ \Delta Z=1.5 \, ,
\label{eq:dY_dZ}
\end{equation}
following Section 3.1 of \citet{choi2016}. Once Y is computed for a value of Z, we calculate X as $X=1-Y-Z$. For details, see Section 3.1 of \citet{choi2016}.  

\item {\it Rotation rates}: We initialise our stars to begin in solid body rotation, at rates parameterised by the ratio of the surface rotation velocity $v$ to the critical velocity for breakup $v_{\rm crit}$, where $v_{\rm crit}$ at the equator is defined as,
\begin{equation}
v_{\rm crit}=\sqrt{(1 - \Gamma_{\rm{edd}}) \, GM/R^3} \, ,
\label{eq:vcrit}
\end{equation}
where 
\begin{equation}
\Gamma_{\rm{edd}} =\kappa L/(4\pi c G M)\, ,
\label{gamma_Edd}
\end{equation}
is the Eddington factor with $\kappa$, $L$, $c$, $M$ and R being the opacity, luminosity of the star, speed of light, mass and radius of the star respectively, all evaluated at the ZAMS (Zero Age Main-Sequence) following the prescription described in Section 3.5 of \citet{choi2016}. We consider $v \, / \, v_{\rm crit}$ from 0 to 0.6 in steps of 0.1. Theoretical models for massive star formation predict a typical rotation tate $v \, / \, v_{\rm crit} \sim 0.5$ \citep{Rosen12a}, but this is significantly uncertain. We therefore adopt a fiducial rotation rate $v \, / \, v_{\rm crit} = 0.4$ wherever the rotation rate is not explicitly mentioned because $v \, / \, v_{\rm crit} = 0.4$ if frequently used as a standard rotating case in other models of massive stellar evolution; in practice this choice matters little, because we find qualitatively similar results for any $v \, / \, v_{\rm crit} \gtrsim 0.4$.  We caution that, thus far, empirical determinations suggest birth rotation rates are somewhat more modest, although this is subject to very large modelling uncertainties in the effectiveness of spin-down via stellar wind torques \citep{Ramirez-Agudelo13a, Ramirez-Agudelo15a}.
\end{itemize}

We run all calculations using the same spatial and temporal resolution conditions as adopted by \citet{choi2016}. Issues of convergence are discussed in \aref{conv_test}.

\begin{figure*}
\centerline{
\includegraphics[trim={0.5cm 0cm 0cm 0cm}, clip, width=0.98\textwidth]{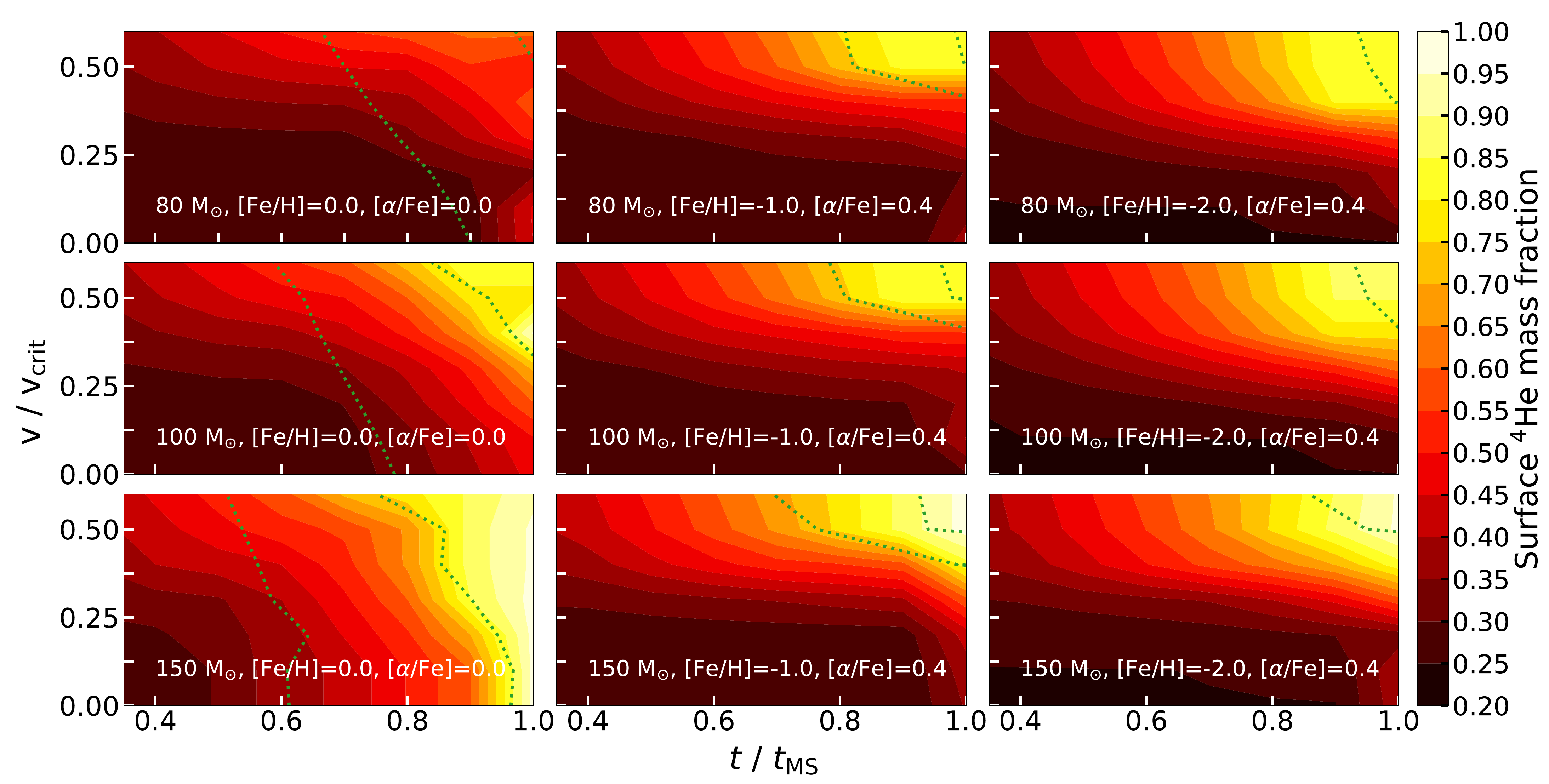}
}

\caption{2D contour map of $^4{\rm He}$ surface mass-fraction as a function of time and rotational velocity for 80 $\mathrm{M}_\odot$ (upper panels), 100 $\mathrm{M}_\odot$ (middle panels) and 150 $\mathrm{M}_\odot$ (bottom panels) stars of three metallicities, $[\rm{Fe}/\rm{H}]=0.0$, $[\alpha/\rm{Fe}]=0.0$ (the leftmost panels); $[\rm{Fe}/\rm{H}]=-1.0$, $[\alpha/\rm{Fe}]=0.4$ (middle panels); $[\rm{Fe}/\rm{H}]=-2.0$, $[\alpha/\rm{Fe}]=0.4$ (the rightmost panels). Note that we have normalised time to the main sequence lifetime $t_{\rm MS}$ for each model, and velocity to the breakup velocity $v_{\rm crit}$. The colourbar shows contours of $^4{\rm He}$ abundance.  The green-dotted lines show the points at which stars have lost 20\% and 50\% of their initial mass.
}
\label{fig:contour}
\end{figure*}

\begin{figure*}
\centerline{
\includegraphics[trim={0.5cm 0cm 0cm 0cm}, clip, width=0.98\textwidth]{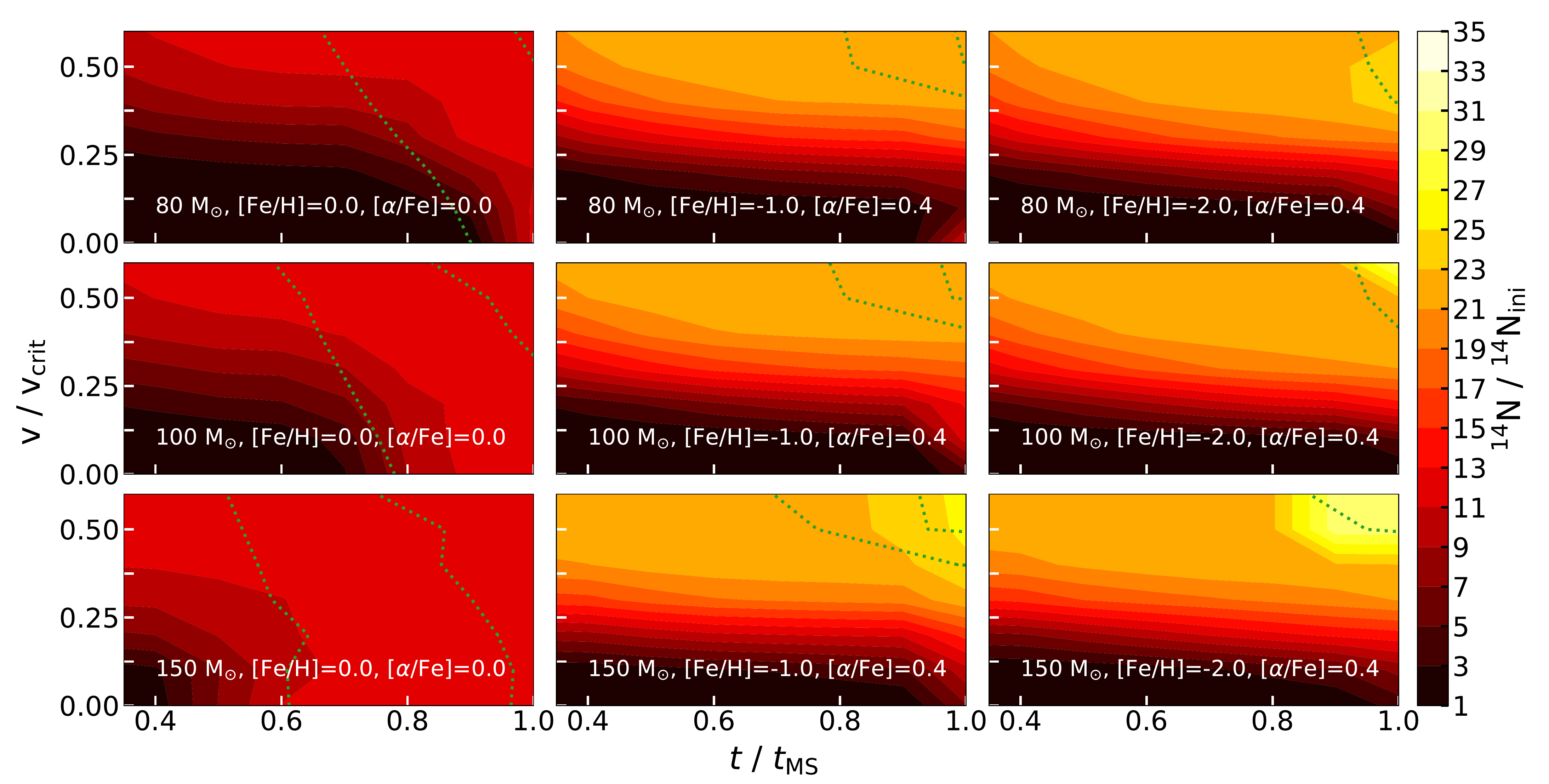}
}

\caption{Same as \autoref{fig:contour}, except here colours indicate $^{14}{\rm N}$ surface mass-fraction normalised by initial $^{14}{\rm N}$ abundance; the initial abundances are $6.73\times 10^{-4}$, $6.97\times 10^{-5}$, and $7.024\times 10^{-6}$ for the $[\mathrm{Fe}/\mathrm{H}]=0.0$, $-1.0$, and $-2.0$, cases, respectively.
}
\label{fig:contour_n14}
\end{figure*}

\begin{figure*}
\centerline{
\includegraphics[trim={0.5cm 0cm 0cm 0cm}, clip, width=0.98\textwidth]{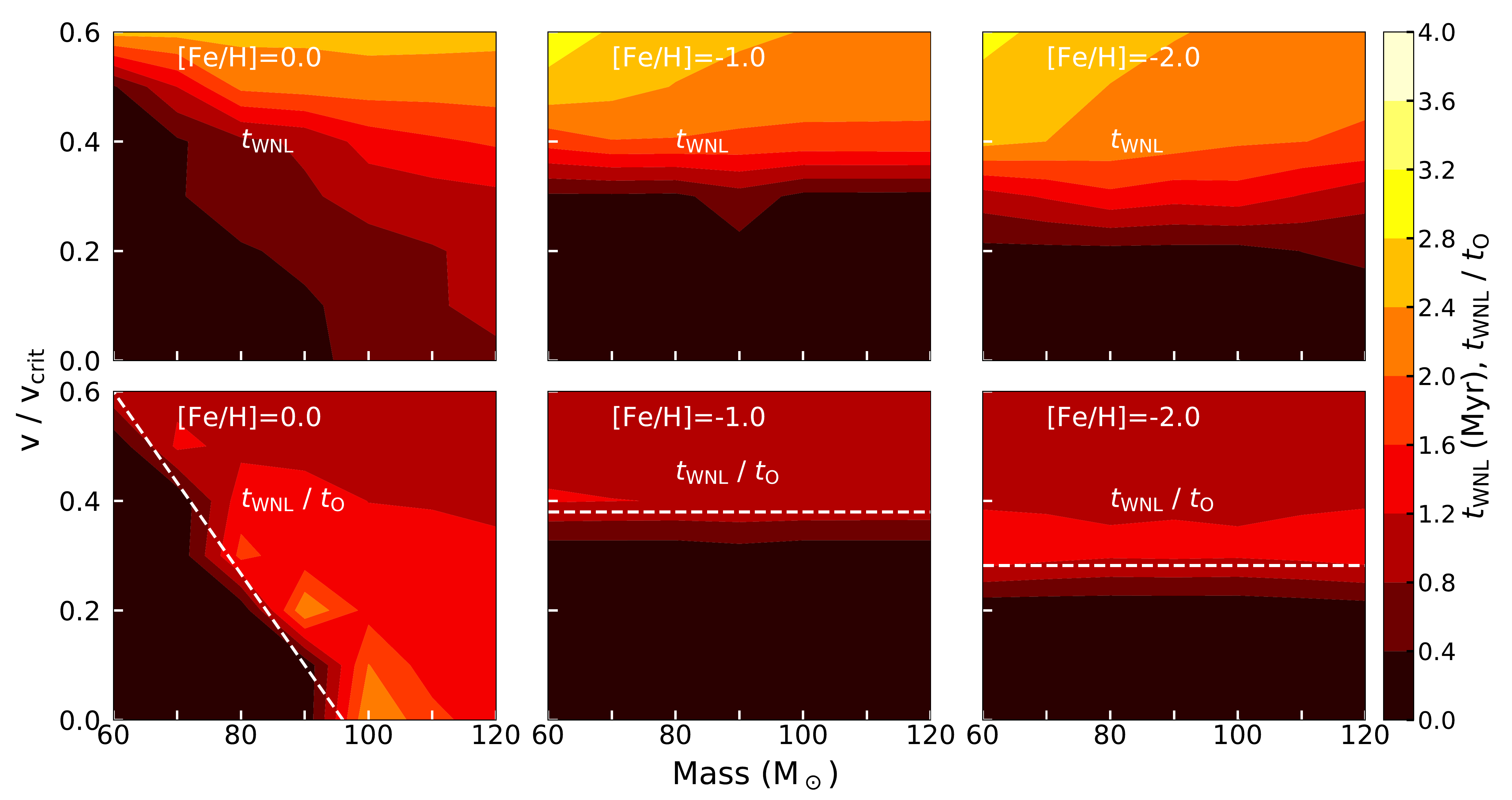}
}

\caption{Contours of t$_{\rm{WNL}}$ (the top panels), the time a star spends in the WNL phase, and   $t_{\rm WNL}/t_{\rm O}$ (the bottom panels), the ratio of the time a star spends in the WNL phase to the time it spends as an O star, as a function of rotation rate and mass for three metallicities, [Fe/H]=0.0 (the leftmost panels), [Fe/H]=-1.0 (middle panels), [Fe/H]=-2.0 (the rightmost panels). The white dashed lines in the bottom panels show $t_{\rm WNL} \approx t_{\rm O}$.}
\label{fig:contour_WNL_O}
\end{figure*}





\section{Results}
\label{results_sec}
The results of our grid of stellar evolution models are described as follows: In
\autoref{param}, we give an overview of stellar surface mass-fractions in our model grid and their implications for the origin of WNL stars using our fiducial choices for all physical parameters. In \autoref{mix}, we discuss the mechanisms by which rotating and non-rotating stars of varying metallicity become WNL stars. In \autoref{param_sensitivity_sec}, we discuss the how our results depend on our parameterisations of uncertain processes such as rotational mixing and massive stellar winds, and discuss the extent to which plausible variations in them might alter our results; in  \autoref{connect_spec_evol}, we discuss the different characteristics of WNE and WNL stars, and the connection between the spectroscopic definition and evolutionary phases of different subtypes of WN-stars. In \autoref{spectra_sec}, we show how our broad range of parameters and mixing mechanisms produces WNL-like surface gravities and temperatures along with WNL surface compositions, and show that synthetic spectra from these stars agree well with observed WNL stars.  

\begin{figure*}
\centerline{
\epsfxsize=1.0\textwidth
\epsfbox{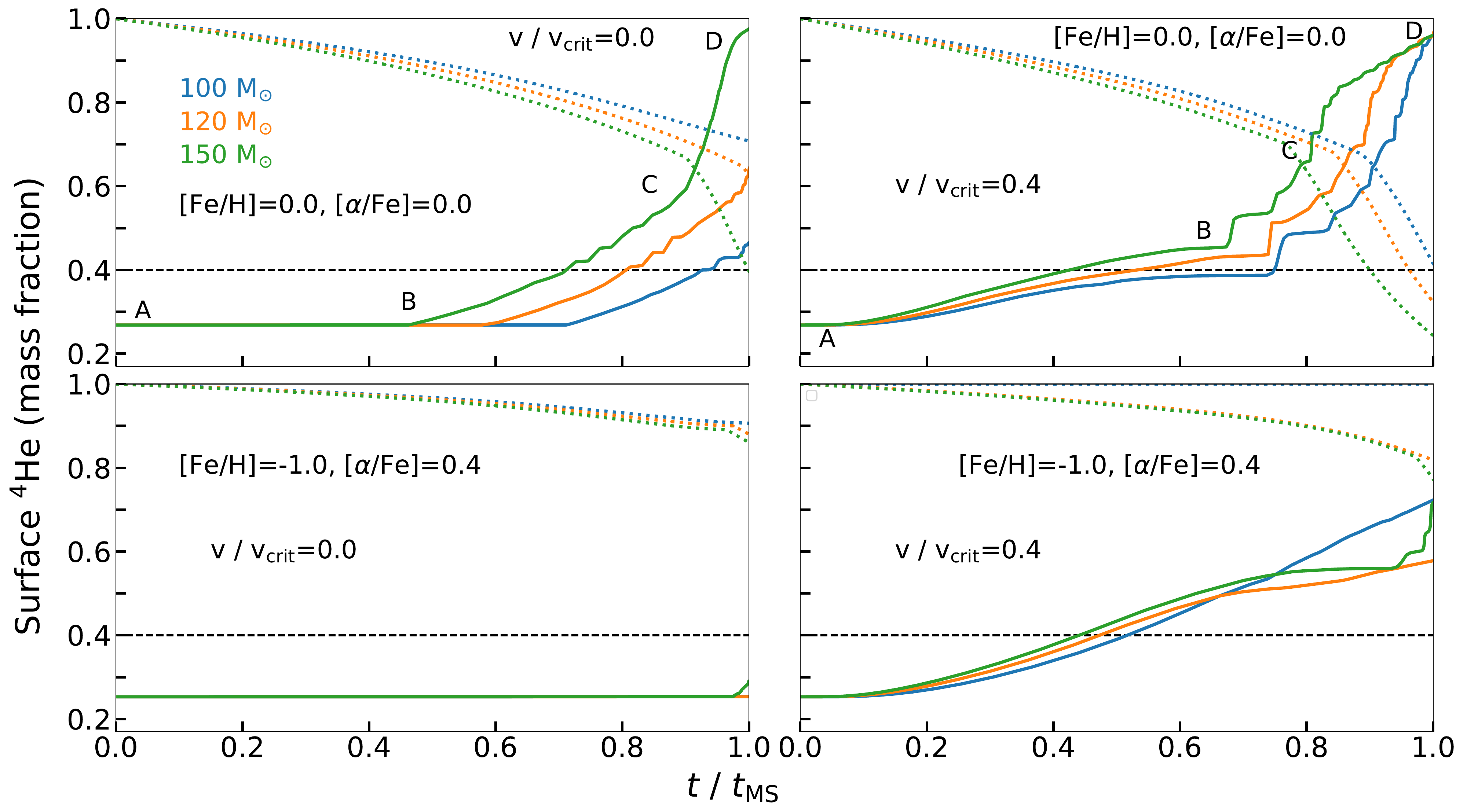}
}
\caption{
Solid lines show surface helium abundance evolution for three different masses (100, 120, 150 $\mathrm{M}_\odot$), two metallicities ($[{\rm Fe}/{\rm H}]=0, -1$), and two rotation rates ($v \, / \, v_{\rm crit}=0, 0.4$), as indicated by the labels on the panels. Dotted lines show stellar mass normalised to initial mass. The dashed black line at a He fraction of 0.4 marks the He abundance traditionally used to delineate the start of the WR phase of stellar evolution \citep[e.g.,][]{meynet2005}.
}

\label{fig:w_wo_rot}
\end{figure*}

\subsection{Overview of parameter space}
\label{param}

\begin{figure}
\centerline{
\epsfxsize=0.5\textwidth
\epsfbox{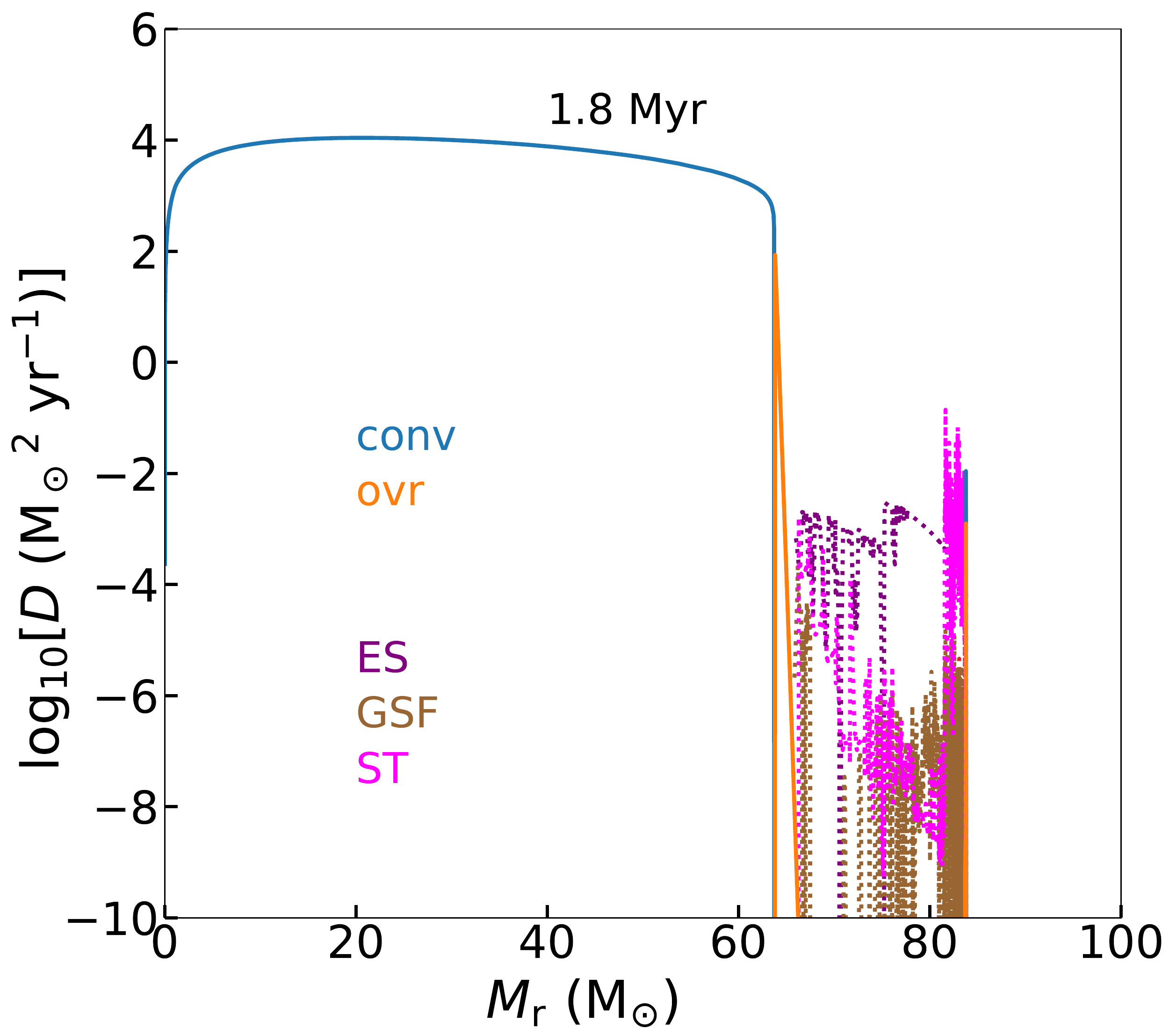}
}

\caption{
Diffusion mixing coefficients as a function of Lagrangian mass coordinate ($M_{\rm r}$) for a $100$ $\mathrm{M}_\odot$ star born with $v \, / \, v_{\rm crit} = 0.4$, $[{\rm Fe}/{\rm H}]=0$, at an age of $1.8$ Myr.
 We show diffusion coefficients of the primary rotational (dotted lines) and non-rotational (solid lines) mixing processes. The non-rotational mixing mechanisms such as convective (conv) and overshoot-convective (ovr) mixings play important roles in mixing and transporting the chemical elements in the convective zones, and similarly, the rotational instabilities such as Eddington-Sweet circulation (ES), Goldreich-Schubert-Fricke instability (GSF), and instabilities due to Spruit Taylor (ST) dynamo have major roles in the radiative zone. 
}
\label{fig:rot}
\end{figure}

We summarise the key results from our grid of MESA models in \autoref{fig:contour} and \autoref{fig:contour_n14}. They show contour plots of the surface mass-fractions of He and N, respectively, as a function of time and rotation rate for a range of initial stellar masses and metallicities. We only show results up to the core H depletion phase, defined as the cessation of core hydrogen burning when the H mass fraction in the core falls below X$_{\rm H}\leq 10^{-6}$. For main sequence stars, there are two distinct enrichment regimes where the surface helium and nitrogen mass fractions are significantly enhanced compared to the star's initial composition: (i) for high metallicity regardless of rotation rate, and (ii) for high rotation rate regardless of metallicity. Both enhancement regimes grow in prominence at higher initial mass.

The figure also shows that the surface He and N enhancements are not being produced by the classical WR process, whereby mass loss removes the stellar envelope and reveals the core. The green-dotted lines in the contour plot show the points by which stars have lost 20\% (the leftmost line) and 50\% (the rightmost line) of their initial mass. Most of the high surface helium regions appear to the left of the 50\% mass-loss contours meaning that the  He surface enhancement is occurring without removal of most of the envelope. For low metallicities, wind loss becomes negligible even with high rotational velocities and thus the 50\% mass-loss contours are absent, yet surface mass-fraction enhancement still occurs.


The levels of surface N enhancement indicated by \autoref{fig:contour_n14} are entirely consistent with the values inferred for WNL stars.
For the most massive stars ($\geq 100$ $\mathrm{M}_\odot$) with high metallicities, we obtain a maximum N-enhancement a factor of $\sim 17$, irrespective of the rotation rate. For our initial $^{14}{\rm N}$ abundance of $6.73\times 10^{-4}$ in the $[{\rm Fe}/{\rm H}]=0$ case, this corresponds to a maximum $^{14}$N surface mass-fraction of $\sim 0.011$. This is within a factor of 2 of the N abundances inferred for WNL stars in the Arches cluster by \citet{figer2002}. For the low metallicity stars, we obtain the nitrogen enhancements by a factor of $\sim 28- 30$ compared to their initial abundances of $\sim 6.97\times 10^{-5}$ and $7.024\times 10^{-6}$ for $[{\rm Fe}/{\rm H}]=-1$, $-2$, respectively.

Our grid also allows to quantify the range of masses, metallicities, and rotation rates that produce a WNL phase. Since WNL stars are distinguished from WNE stars, at least in part, by the presence of hydrogen features, for the purposes of this paper we define the WNL phase of a star's life to be when the He surface mass-fraction is 40\%$\leq Y_{\rm{He}}\leq 90$\% (\autoref{tab:def_tab}). We verify below that the stars we select by this criterion in fact have spectra that provide a reasonable match to observed WNL spectra. For comparison, we also define O-stars as the evolutionary phase when the surface H-abundance  $X_{\rm{H}}\geq 60$\% and the star is still on the MS. The WNL and O-star lifetimes are denoted as $t_{\rm{WNL}}$ and $t_{\rm{O}}$ respectively.

We show contours of $t_{\rm{WNL}}$ and $t_{\rm{WNL}}/t_{\rm{O}}$  as a function of mass and rotation rate for three metallicities in the top and bottom panels of  \autoref{fig:contour_WNL_O} respectively. We find, in the top panels of \autoref{fig:contour_WNL_O}, that the lifetime of the WNL phase can be as long as $\sim 2.6 - 3.2$ Myr, similar to the average lifetime of the most-massive O-stars for solar and sub-solar metallicities down to [Fe/H]=-2.0. 

\begin{table}
\begin{tabular}{cc}
\hline\hline
$v \, / \, v_{\rm crit}$ & Mass at which $t_{\rm{WNL}}$/$t_{\rm O}=1$\\
& [$\mathrm{M}_\odot$] \\
\hline
0.6 & 60 \\
0.5 & 70\\
0.4 & 72\\
0.3 & 78\\
0.2 & 85\\
0.1 & 90\\
0.0 & 96\\

\hline\hline
\end{tabular}
\caption{
\label{tab:WNL_tab}
Mass at which $t_{\rm{WNL}}=t_{\rm{O}}$ for different values of $v \, / \, v_{\rm crit}$ at solar metallicity. $t_{\rm{WNL}}$ and $t_{\rm{O}}$ are the times star spends in the WNL-phase and as an O-star respectively.
}
\end{table}

The white dashed lines in the bottom panels of \autoref{fig:contour_WNL_O} indicate $t_{\rm WNL} \approx t_{\rm O}$, so that stars with masses and rotation rates that place them above the line spend more of their lives as WNLs than as O stars. For solar metallicity, the transition line from O-star dominated to WNL-star dominated depends on both mass and rotation rate, whereas for low metallicities, the transition lines are dependent only on rotation rates and are independent of stellar masses down to the lowest mass models, $60$ $\mathrm{M}_\odot$, in our grid. For moderately rotating solar-metallicity stars ($v \, / \, v_{\rm crit}=0.4$), the ``transition" mass between stars that spend most of their lives as O stars and those that spend most of their lives as WNL stars is $\sim 75- 80$ $\mathrm{M}_\odot$. We provide this value for other rotation rates in \autoref{tab:WNL_tab}, and a fit to the white-dashed line in the leftmost bottom panel  gives $M_{t_{\rm WNL}=t_{\rm O}} = (95 - 58.3$  $v \, / \, v_{\rm crit})\mathrm{M}_\odot$. If we assume that most massive stars are born rotating at $v \, / \, v_{\rm crit}=0.5$ following \citet{Rosen12a}, then for solar metallicity we find all the stars more massive than 70 $\mathrm{M}_\odot$ will spend $\sim 20$\% more time in the WNL-phase compared to O-star phase during its lifetime. Assuming Salpeter IMF (dN/dM$\propto$ M$^{-2.35}$), this means that on average there should be 1 star that will experience a WNL phase at some point during its lifetime in any cluster with a mass $\gtrsim 10^4$ $\mathrm{M}_\odot$.  
For sub-solar metallicity stars, the transition from mostly O to mostly WNL is driven mainly by rotation, and occurs at $v \, / \, v_{\rm crit} \approx 0.4$ and $0.3$ for [Fe/H] of $-1$ and $-2$, respectively. Assuming that low-metallicity massive stars are mostly born as moderately fast rotators, similar to more metal rich massive stars (as suggested by theoretical models for massive star rotation -- \citealt{Rosen12a}), this means that WNL stars should become increasingly common in low metallicity populations. For sub-solar stars with $v \, / \, v_{\rm crit}\geqslant 0.4$,  the cluster mass at which we expect there to be at least one star that goes through a WNL phase decreases slightly to $\approx 8000$ $\mathrm{M}_\odot$.

\begin{figure}
\centerline{
\includegraphics[width=0.5\textwidth]{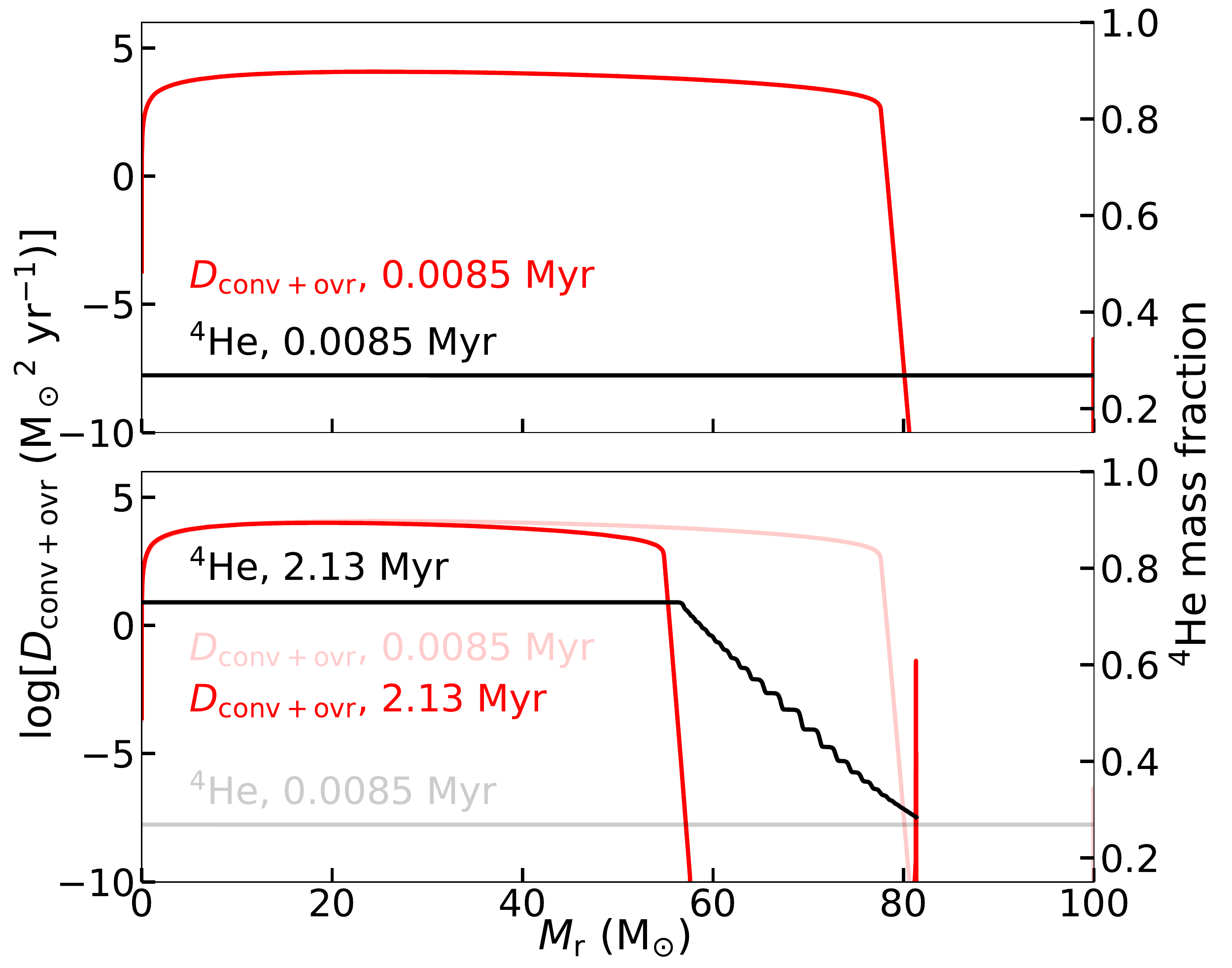}
}

\caption{Diffusion coefficients of non-rotational mixing mechanisms (convective+overshoot mixing) and the helium mass fraction as a function of Lagrangian mass coordinate ($M_{\rm r}$) for a $100$ $\mathrm{M}_\odot$, $[{\rm Fe}/{\rm H}] = 0.0$ star at two times: early in the evolution ($0.0085$ Myr, top panel) and at the start of phase `BC' ($2.13$ Myr, bottom panel). In the bottom panel, the faded lines match those plotted in the top panel, and are provided for comparison.
}
\label{fig:overlap_conv}
\end{figure}

\subsection{Physical mechanisms}
\label{mix}
Having established that there appear to be two distinct regimes of surface He and N enhancement, we next investigate the physical mechanisms responsible. To aid in this discussion, we focus on four points in our model grid: rotation rates of $v \, / \, v_{\rm crit} = 0$ and $0.4$, and initial abundances $[\mathrm{Fe}/\mathrm{H}] = 0$ ($[\alpha/\mathrm{Fe}] = 0$) and $[\mathrm{Fe}/\mathrm{H}] = -1$ ($[\alpha/\mathrm{Fe}] = 0.4$). All of these cases except $v \, / \, v_{\rm crit} = 0$, $[\mathrm{Fe}/\mathrm{H}] = -1$ show significant surface He and N enhancement for stars still on the main sequence, before significant mass loss occurs. For each combination of rotation rate and metallicity, we plot the surface He mass fraction as a function of time for a sample of initial stellar masses in \autoref{fig:w_wo_rot}. The figure reveals that there are three distinct phases  during which the surface He abundance builds up, which we label `AB', `BC', and `CD' in the plot. We now explore the physical origins of each phase.

\subsubsection{Rotational mixing}
\label{rot}
The first phase of He enhancement visible in \autoref{fig:w_wo_rot}, denoted `AB', occurs only for the rotating case, and begins almost immediately. This mechanism enhances the surface He abundance to 40\% - 50\% by ages $\approx 2.5$ Myr for stars of $100$ $\mathrm{M}_\odot$ or more.
A few important mixing mechanisms work simultaneously in rotating stars. As an example, we show the radial profiles of the diffusion coefficients of the primary mixing mechanisms  for a 100 $\mathrm{M}_\odot$ star with $v \, / \, v_{\rm crit} = 0.4$ and $[{\rm Fe}/{\rm H}=0]$ at an age of 1.8 Myr, during phase `AB', in \autoref{fig:rot}. The figure shows that the most important non-rotational mixing mechanisms are convective and overshoot-convective mixing. For rotating stars, the two convective zones (inner core and outer shell) are connected by three dominant rotational transport mechanisms: meridional circulation (Eddington-Sweet circulation), Goldreich-Schubert Fricke instability, and Spruit dynamo mixing. In non-rotating stars during phase `AB', there are no diffusion mechanisms for the transport of chemical elements from  the inner convective core to the outer convective shell, and therefore no surface enhancement occurs, as shown in the topmost left panel of \autoref{fig:w_wo_rot}. The effect of rotational mixing of chemical elements has previously been studied by a number of authors \citep[e.g.,][]{maeder2005, heger2000, meynet2005, meynet2000, crowther2007}.


\subsubsection{Exposing ``fossil" convective cores}
\label{fossil_conv}
The second phase of helium enrichment is denoted as `BC' in \autoref{fig:w_wo_rot}. This mechanism begins at $\sim 1.2- 2.2$ Myr and ends after $\sim 2.5- 3$ Myr, and enhances the surface He-mass fraction to $\sim 50$\% and $60- 65$\% for the 100 $\mathrm{M}_\odot$ non-rotating and rotating stars respectively.
This phase of surface He enhancement in non-rotating stars is a result of two pieces of physics acting together:

(i) When very massive stars are young, their inner convective zones are very big. In \autoref{fig:overlap_conv}, we show the radial profiles of the He mass fraction and non-rotational diffusion coefficient for a 100 $\mathrm{M}_\odot$ star at two times: immediately after the star forms ($\approx 0.01$ Myr), and at the start of phase `BC' ($\approx 2.1$ Myr). The profiles of N are similar to those of He, but we omit them to avoid clutter. At the earlier time, the inner convective zone goes out to 80 $\mathrm{M}_\odot$ as shown in the top panel of \autoref{fig:overlap_conv}. Consequently, nucleosynthetic products get mixed out to large masses/radii. As the star evolves, the convective zone shrinks and by 2.13 Myr, during phase `BC', it ends inside the $60$ $\mathrm{M}_\odot$ shell, as shown in the bottom panel of \autoref{fig:overlap_conv}. The gradual retreat of the convective core leaves behind a smooth gradient in He abundance that is in effect a record of how long a particular Lagrangian mass shell spent inside the inner convective zone. The 80 $\mathrm{M}_\odot$ shell shows almost no He enhancement, because that shell falls out of the convective zone only a $\mathrm{few}\times 10$ kyr into the star's life. By contrast, the $60$ $\mathrm{M}_\odot$ shell remains inside the convective core for $\approx 2$ Myr, by which time the helium abundance has increased to $\approx 70\%$. 

\begin{figure}
\centerline{
\includegraphics[width=0.5\textwidth]{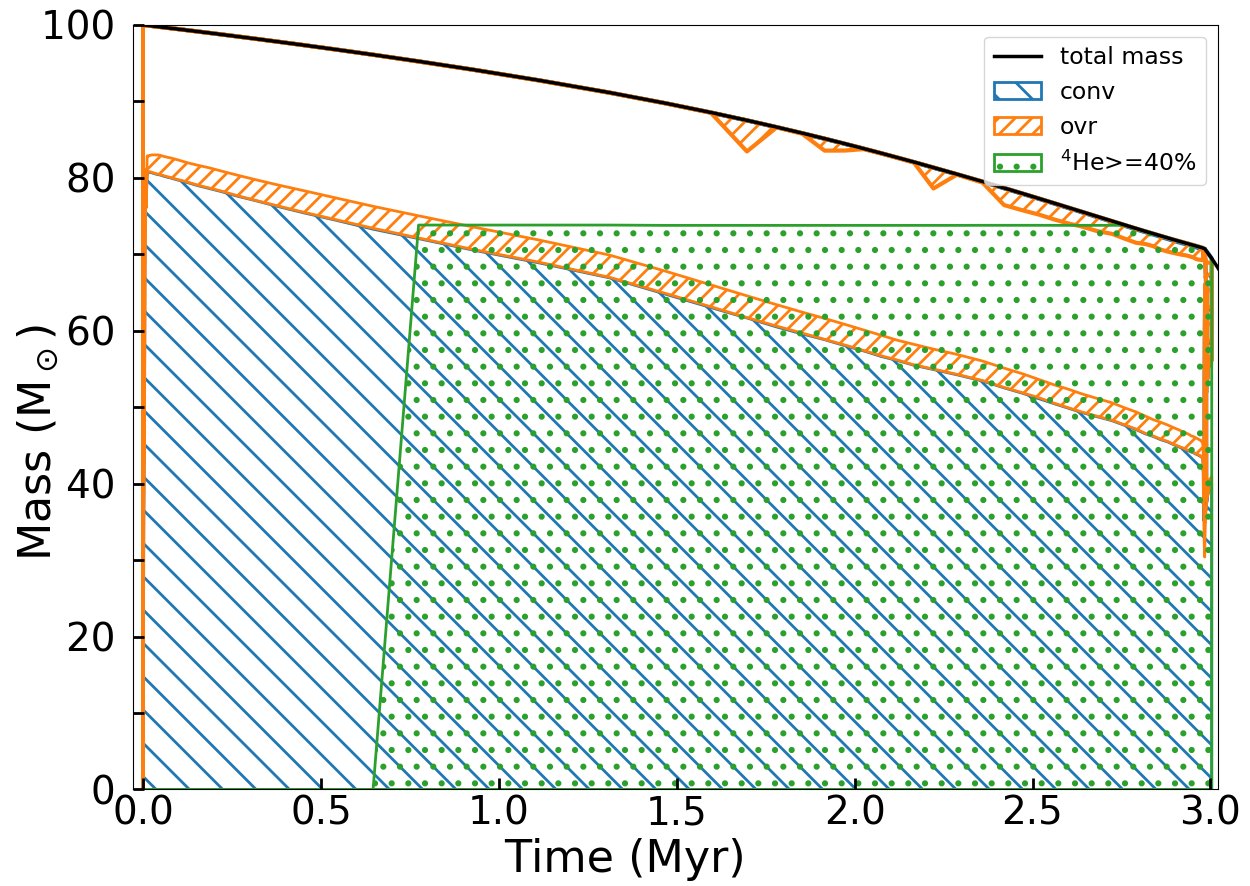}
}

\caption{Kippenhahn diagram of a 100 $\mathrm{M}_\odot$, solar metallicity, non-rotating star. The black solid line shows the time evolution of stellar mass. The blue, orange, and the green hatched regions indicate Lagrangian cells representing the regions of convection, overshoot convection, and He mass fraction $\ge 40$\%, respectively.
}
\label{fig:kippenhahn}
\end{figure}

\begin{figure}
\centerline{
\epsfxsize=0.5\textwidth
\epsfbox{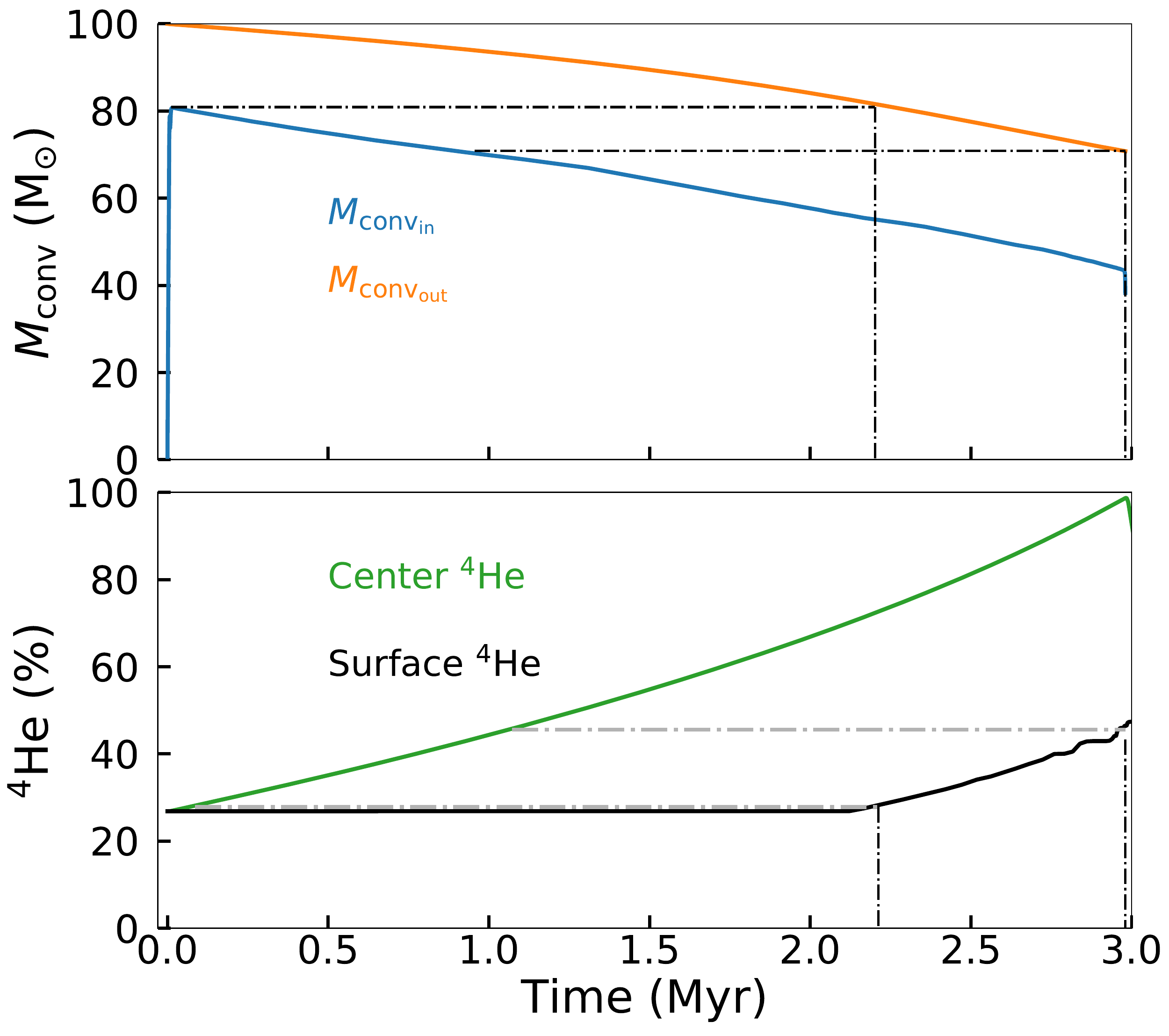}
}

\caption{
{\it Top panel:} Lagrangian mass coordinate marking the outer edge of the inner and outer convective zones as a function of stellar age for a 100 $\mathrm{M}_\odot$, non-rotating, solar metallicity star. The left vertical black dashed-dotted line marks the start of phase `BC', corresponding to the time at which the outer convective zone of the star (which is nearly at its surface) reaches the Lagrangian mass shell that marked the outer edge of the convective core when the star first formed. The right vertical black dashed-dotted line marks the time when core H depletion is over. The horizontal black dashed-dotted lines show the ``fossil"-convective core masses of previous times, which match with the Lagrangian mass cells of the present epoch in outer convective zones.
 \textit{Bottom panel:} He mass fraction in the stellar core and at the stellar surface. The values of the surface He-abundances of the present epoch matching the central He-abundances of the ``fossil" epochs are shown by the horizontal grey dashed-dotted lines.  
}
\label{fig:conv_in_out}
\end{figure}

\begin{figure}
\centerline{
\epsfxsize=0.5\textwidth
\epsfbox{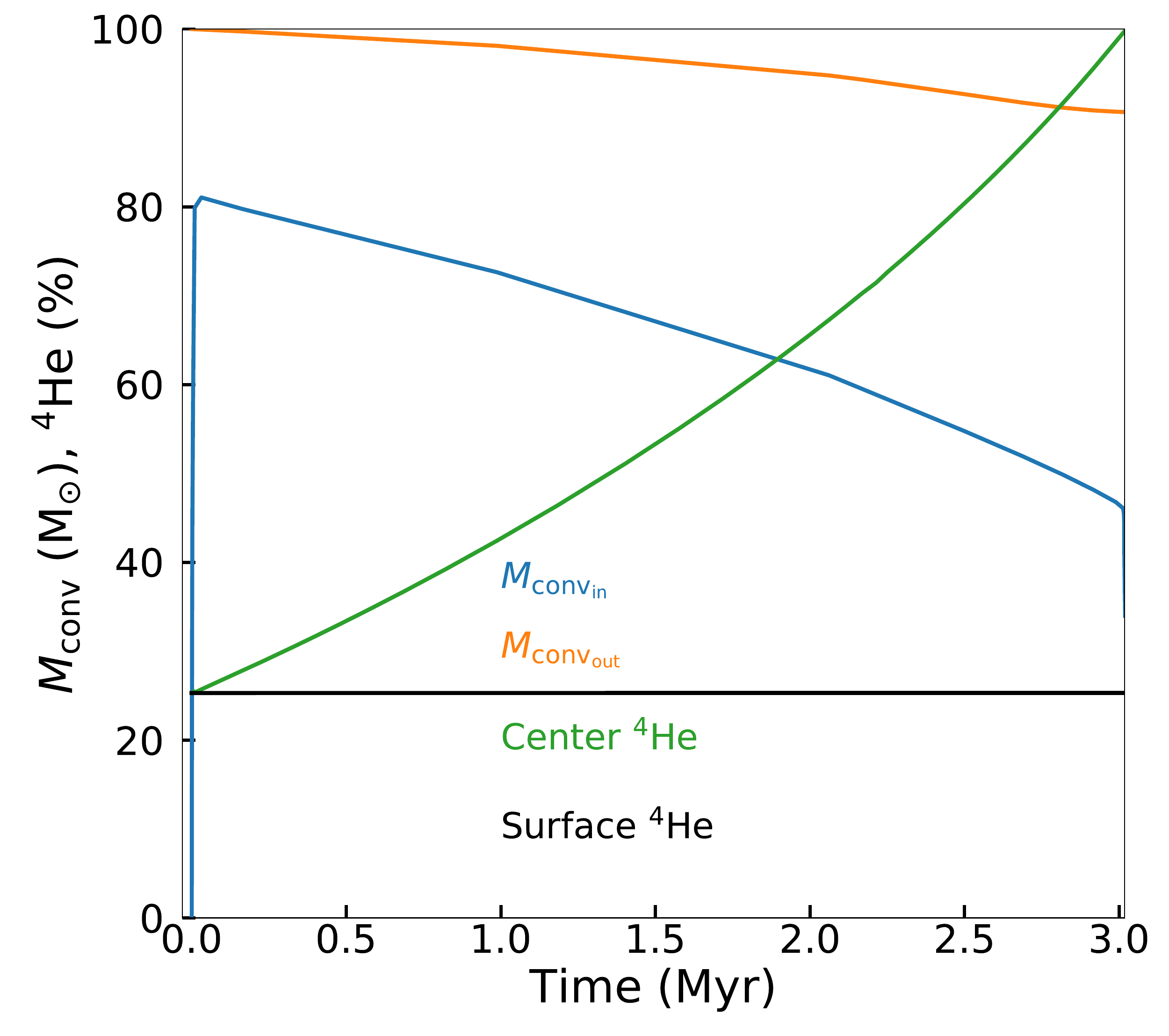}
}
\caption{Same as \autoref{fig:conv_in_out}, except here we show the metal-poor ([$\mathrm {Fe}$/$\mathrm H$]=-1.0, [$\alpha$/$\mathrm {Fe}$]=0.4) non-rotating 100 $\mathrm M_{\odot}$ sample star. 
}
\label{fig:conv_low_feh}
\end{figure}

(ii) The other mechanism that interacts with (i) is mass loss on the main sequence. Even before the classical WR phase, O stars lose some mass through winds. Because the convective zone was once very large in these very massive stars, it does not take much mass loss to reach where the convective zone once was. The more massive the star is, the further out its initial convective core goes, and the less mass loss is needed to reach it. In effect mass loss is revealing a ``fossil''-convective core: parts of the star that were once part of the convective core but are not any more. 
We can see this effect at work in the bottom panel of \autoref{fig:overlap_conv}, where we plot the He mass-fraction and convective core locations at an age of $2.1$ Myr (black and red lines), along with the ``fossil'' values of these quantities at an age of $0.01$ Myr (grey and faded red lines). At $2.1$ Myr, the star has lost the outer $\approx 20$ $\mathrm{M}_\odot$ of its mass through O star winds, and thus its outer surface now lies at the same Lagrangian mass shell that marked the outer edge of the convective core when the star was $\approx 0.01$ Myr old. Consequently, its surface He abundance matches the He abundance in that shell that prevailed at the time when the convective core retreated past it. This is why $2.1$ Myr marks the start of phase `BC' for this star: it is the age at which mass loss has finally eaten down to the location where the convective core ended shortly after the star formed.


Once mass loss has reached the outer edge of the fossil convective core, continuing mass loss causes the stellar surface to dive deeper into the core, revealing progressively more He- and N-enhanced material. The subsequent evolution of the surface He and N abundances is then dictated by the combination of the radial profile of He and N set up by retreat of the convective zone and the rate at which mass loss reveals the fossil enhancements left behind. 
To help visualise the phenomenon, we consider the example of a 100 $\mathrm{M}_\odot$, non-rotating, solar metallicity star. We plot a Kippenhahn diagram for this star in \autoref{fig:kippenhahn}; to accompany this, we plot the Lagrangian mass coordinates marking the outer edges of the inner and outer convective zones for this star in the top panel of \autoref{fig:conv_in_out}, and the time evolution of the central and surface He abundances for this star in the bottom panel. At a very early times, these figures show that the inner convective region extends to almost $\sim 80$ $\mathrm{M}_\odot$, and all of this mass begins to enrich in He. Mass loss exposes this Lagrangian shell at $\sim 2.1- 2.2$ Myr, indicated by the dashed vertical lines in \autoref{fig:conv_in_out}, and this marks the start of phase `BC'. From that point forward, as shown in the lower panel \autoref{fig:conv_in_out}, the surface He fraction simply tracks the core He fraction at the time when the current outer shell ceased to be part of the convective core. Thus for example, \autoref{fig:kippenhahn} shows that the core He fraction reaches $40\%$ after $\approx 0.8$ Myr of evolution, at which point the convective zone covers the inner $\approx 75$ $\mathrm{M}_\odot$ of material. This shell becomes exposed by mass loss at $\approx 2.7$ Myr of evolution, at which point the surface He fraction becomes 40\%, and we would classify the star as WNL.

This explanation makes is clear why surface He and N enhancements only occur for non-rotating stars if they are relatively metal rich: despite low-metallicity massive stars' very large convective cores, they lose mass so slowly via O star winds that their cores are not revealed before core hydrogen is exhausted and the star leaves the main sequence. We also show the time evolution of the Lagrangian mass coordinates marking the outer edge of the inner convective zone and inner edge of the outer convective zone, and central and surface He mass-fraction for a metal-poor ([$\mathrm {Fe}$/$\mathrm{H}$]=-1.0) non-rotating 100 $\mathrm{M}_{\odot}$ sample star in \autoref{fig:conv_low_feh}, similar to \autoref{fig:conv_in_out}. We find that the outer convective zone never becomes part of the inner convective zone of previous times until the end of the core H depletion phase in \autoref{fig:conv_low_feh}.

The phenomenon of fossil convective core exposure is most easily seen for non-rotating stars, but it also occurs in metal-rich rotating stars such as those shown in the upper right panel of \autoref{fig:w_wo_rot}. In this case, as for the non-rotating counterparts, there is a clear jump in the rate at which the surface He abundance rises at $\approx 2$ Myr of evolution, denoted as `B' in the figure. No such jump is apparent in the low-metallicity case shown in the lower right panel.

The fossil convective core phenomenon we have discovered provides a natural explanation for the slowly- / non-rotating He- and N-enhanced stars observed by \citet{herrero2000} and \citet{vink2017}. \citet{meynet2000} hypothesised that these stars might have been born as rapid rotators that experienced rotationally-driven mixing during the MS but then suffered catastrophic angular momentum loss during the WR phase to become the slow-rotators. Here we have shown that there is no need for so complex a scenario: even slowly-rotating very massive stars will reveal He and N surface enhancements as a result of fossil convective core exposure.

\subsubsection{``Classical" WR winds mass-loss}
\label{WR}

The third phase of enhancement is denoted as `CD' in \autoref{fig:w_wo_rot}. During this phase, the surface He abundance increases to $90-95$\% towards the end of or shortly after the core H depletion for all stars except those that are both non-rotating and metal-poor. 
This is the classical WR phase, caused by very rapid mass loss revealing the current (as opposed to fossil) convective core. The transition between phases `BC' and `CD' is caused by a sharp increase in the mass loss rate, visible as the sharp change in the slopes of the dotted lines in \autoref{fig:w_wo_rot}. This change in mass loss rate is caused by the stellar surface approaching the Eddington limit, and has been discussed extensively in the literature \citep[e.g., see the review by][]{crowther2007}. We shall not discuss it in detail, beyond noting that since the opacity of the atmosphere depends on metallicity and the effective Eddington limit depends on the degree of rotational support, it is not surprising that our models show that this phase occurs for either metal-rich or rapidly-rotating stars, but is absent in stars that are neither, as shown in the lower left panel of \autoref{fig:w_wo_rot}.

\begin{figure}
\centerline{
\epsfxsize=0.5\textwidth
\epsfbox{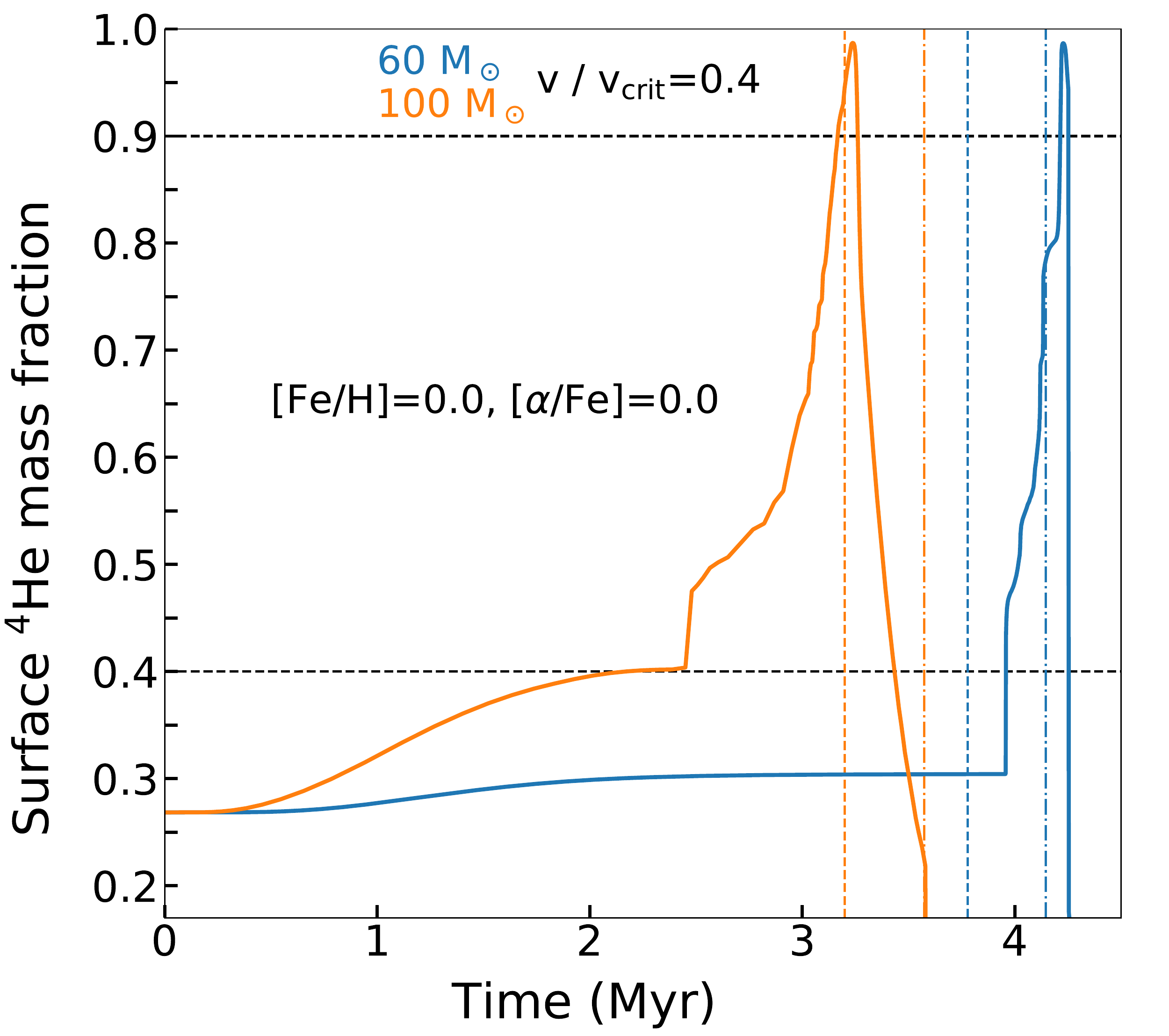}
}

\caption{Time evolution of the surface He abundances for 60 and 100 $\mathrm{M}_\odot$ stars  with $v \, / \, v_{\rm crit}=0.4$ and solar metallicity. The dashed and dashed-dotted vertical lines represent the MS-lifetimes and He-core burning lifetimes respectively. The black dashed horizontal lines represent He surface mass-fractions of 40 and 90\%, respectively. Stars above the top dashed line have He and N abundances that would lead to them being classified as WNE, while those between the two dashed lines have abundances that would be classified as WNL.}
\label{fig:example_WNE_WNL}
\end{figure}

\begin{figure*}
\centerline{
\includegraphics[trim={0.5cm 0cm 0cm 0cm}, clip, width=0.98\textwidth]{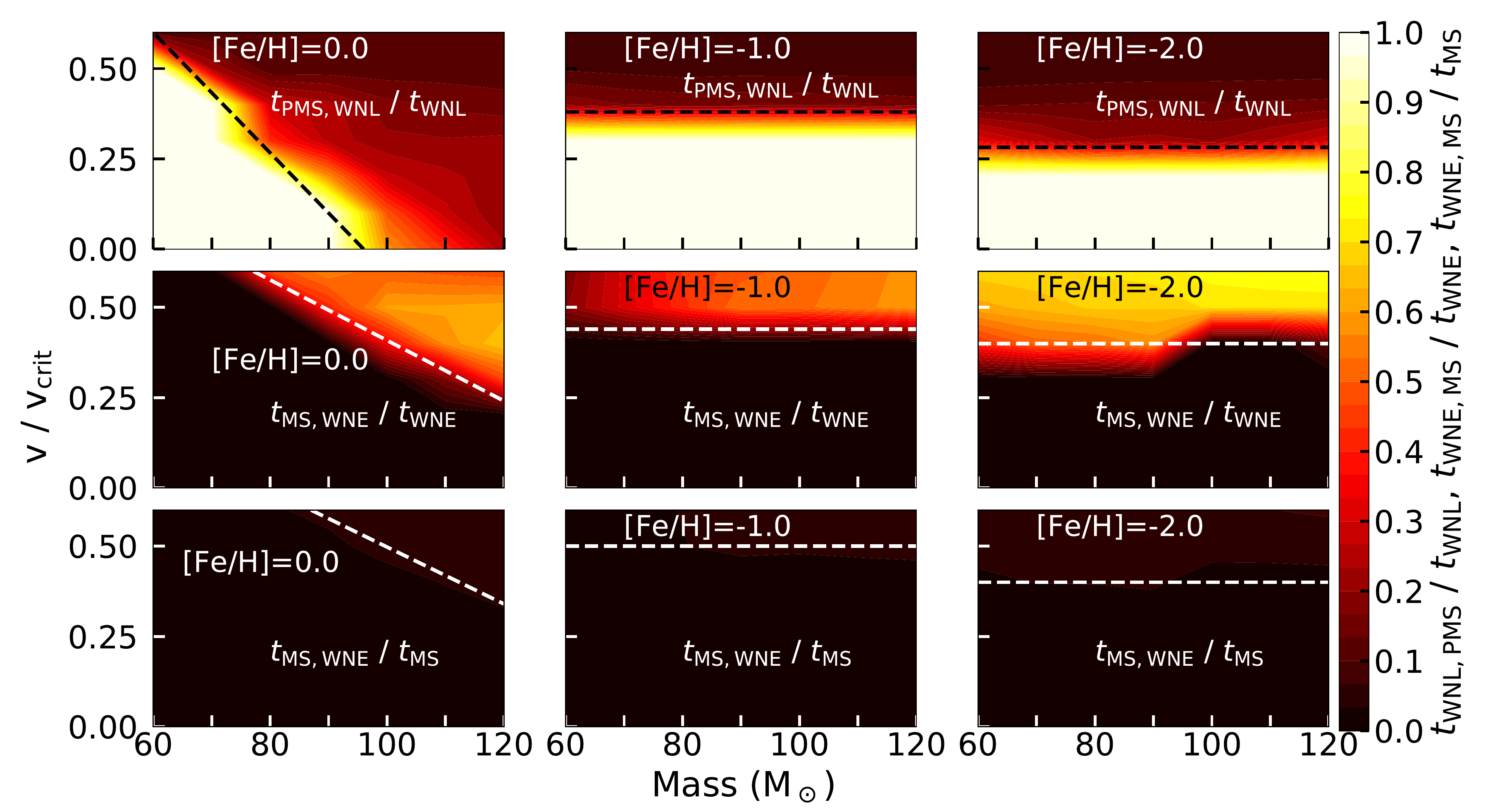}
}

\caption{Contours of t$_{\rm{WNL,PMS}}$/t$_{\rm{WNL}}$ (the top panels), t$_{\rm{MS, WNE}}$/t$_{\rm{WNE}}$ (the middle horizontal panels), and t$_{\rm{MS, WNE}}$/t$_{\rm{MS}}$ (the bottom panels) as a function of mass and rotation rate for three metallicities, [Fe/H]=0.0 (the leftmost panels), [Fe/H]=-1.0 (the middle vertical panels), and [Fe/H]=-2.0 (the rightmost panels), where t$_{\rm{WNL,PMS}}$ and t$_{\rm{WNE,MS}}$ are defined as the time spent by a star in the post-MS and on the MS as WNL and WNE star respectively. The zero values in the middle (horizontal) and bottom panels corresponds to a star that either becomes a WNE only after the MS, or never does.
 The dashed lines in the top panels divide the regions when a star shows WNL features 100\% and $10- 35$\% on the post-MS phase. The white-dashed lines in the horizon middle panels represent the stars showing $\geq 30$\% of the WNE-lifetime on the MS falling above these lines. The white-dashed lines in the bottom most panels represent the stars having $\geq 10- 15$\% of the MS lifetime as a WNE phase falling above these lines.  
}
\label{fig:pms_WNL_ms_WNE}
\end{figure*}
\subsection{Uncertainties arising from rotation, overshoot, and mass loss}
\label{param_sensitivity_sec}

All the results presented thus far are for our fiducial parameter choices, which match those recommended by \citet{heger2005} (for rotational mixing) and \citet{paxton2013} and \citet{choi2016} (for overshoot convection and mass loss). These parameters have been tuned to reproduce observations of massive stars, and thus are empirically motivated. It is, however,  important to understand how varying these parameters might affect our results. We provide full details on the experiments we have carried out to explore this dependence in \aref{param_sensitivity} and \aref{conv_test}, and here simply summarise the main conclusions.

\begin{itemize}
\item \textit{Rotational mixing:} varying the values of the coefficients $D$ and $\gamma$ describing rotation-induced diffusion directly affects the amount of He enrichment that occurs during phase `AB': increasing the diffusion coefficient by a factor of $2-4$ causes the WNL phase to start a factor of $\sim 2$ earlier in time than for a our fiducial value. Reducing the coefficient by a similar factor delays the onset of the WNL phase by $\sim 10\%$ in metal-rich stars, where mass loss during phase `BC' produces a WNL phase even in the absence of rotation, and by a larger amount in metal-poor stars, which only undergo rotational enrichment. From the standpoint of populations, as illustrated in \autoref{fig:contour} and \autoref{fig:contour_n14}, the main effect of varying the rotational mixing coefficients is to increase or decrease the prominence of the horizontal feature that causes early enrichment for $v \, / \, v_{\rm crit} \gtrsim 0.3$. We also find that changes in the spatial resolution are degenerate with variations in the diffusion coefficients, suggesting that numerical mixing is non-negligible, and that the diffusion coefficients must be calibrated to a value that is resolution-dependent.
\item \textit{Convective overshoot:} overshoot mixing proves to be relatively unimportant for our primary findings. If we disable it, the effects are significant only for non-rotating stars, and for them the main effect of turning off overshoot mixing completely is only a slight delay in the onset of the WNL phase.
\item {\it Mass loss:} Varying the mass loss rate upwards or downwards by a factor of 3 significantly affects the occurrence of the WNL phase for non-rotating stars because it alters the exposure of ``fossil"-convective cores. 
Increasing mass loss rates by a factor of 3 causes all solar metallicity stars with $M > 100 \mathrm{M}_\odot$ to spend the majority of their main sequence lives as WNLs even if they are non-rotating, and causes WNLs to begin appearing even among non-rotating stars at [Fe/H]$=-1$. Conversely, decreasing the mass loss rate by a factor of $3$ guarantees that no non-rotating stars become WNLs even at solar metallicity. For rotating stars the sensitivity to the mass loss rate is much less, and the frequency of WNLs is nearly unchanged as the mass loss prescription is varied. Instead, the main effect of increasing or reducing the mass loss rate is to change the maximum level of He enrichment that stars reach before the core H depletion, with higher mass loss rates generally producing peak surface He abundances. Thus in terms of \autoref{fig:contour} and \autoref{fig:contour_n14}, changing the mass loss prescription changes the prominence of the enrichment region in the lower-right portion of the left column of contour plots, at low $v \, / \, v_{\rm crit}$ and high metallicity. However, we remind readers that we have treated our uncertainties in mass loss as a simple scaling of the overall loss rate, independent of stellar properties; a more complex scaling, for example one that depends on effective Eddington ratio, might produce more complex changes.
\end{itemize}

\subsection{Connection between spectroscopic definition and evolutionary phase for WN-subtypes}
\label{connect_spec_evol}

Having described the different mechanisms for the production of WNL-stars and their dependence on parameter sensitivity, in this section we attempt to connect the spectroscopic definitions of the two WN-subtypes (WNE, WNL) with  evolutionary phase. Several authors have proposed a connection between these surface mass-fractions and the evolutionary phase of the stellar core as discussed in \autoref{intro}.

Our model grid enables us to test these proposed identifications.  Unfortunately, we find that there is no direct one-to-one connection between surface mass-fractions and evolutionary phases of WN-subtypes. For example, we show the evolution of surface He abundances for 60 and 100 $\mathrm{M}_\odot$ stars with solar metallicity and a typical rotation rate ($v \, / \, v_{\rm crit}=0.4$) in \autoref{fig:example_WNE_WNL}. We find that both example stars stars exhibit WNL-like surface mass-fractions both during the MS, H core-burning phase and the post-MS, He core-burning phase. Also, the 100 $\mathrm{M}_\odot$ star shows the WNE-like surface He abundance for a brief period immediately preceding core H depletion. Thus for these two example stars, there is no clear mapping between evolutionary phase and surface mass-fraction.

To illustrate this point over a broader parameter space, we define $t_{\rm PMS,WNL}$ as the amount of time for which a star is post-MS (i.e., has ceased core H burning) but has WNL-like surface mass-fractions. We similarly define $t_{\rm MS, WNE}$ as the duration for which a star on the MS has WNE-like abundances. If there were a clean mapping between core evolutionary phase and surface mass-fractions, we would expect both $t_{\rm PMS,WNL}/t_{\rm WNL} \ll 1$ and $t_{\rm MS, WNE}/t_{\rm WNE} \ll 1$ or $t_{\rm MS, WNE}/t_{\rm MS} \sim 0$, where $t_{\rm WNE}$ is defined, in analogy to $t_{\rm WNL}$, as the amount of time for which a given star has WNE-like surface mass-fractions regardless of its core evolutionary state. We plot contours of these ratios as a function of mass and rotation rate for three metallicities ([Fe/H]=0.0, -1.0, -2.0) in \autoref{fig:pms_WNL_ms_WNE}. All the stars falling below the black-dashed lines in the top panels of \autoref{fig:pms_WNL_ms_WNE} spend 100\% of their WNL-phase on the post-MS, and even the stars above these lines spend a considerable fraction ($10- 30$\%) of their time as WNL stars on the post-MS. The white-dashed lines in the horizontal middle and bottom-most panels of \autoref{fig:pms_WNL_ms_WNE} show that the converse holds for WNE stars, i.e., a significant fraction of the WNE phase ($\geqslant$ 35 \% as shown the horizontal middle panels) occurs while stars are still on the MS,  and a significant fraction of the MS lifetime, $\sim$ 5$- 10$ \% as shown in the bottom-most panels,  is spent with WNE-like surface properties. 
Therefore, we conclude that there is no obvious one-to-one connection between the surface mass-fractions of He (and N, which shows enhancements qualitatively similar to those of He), and thus with the spectroscopically-defined WNL and WNE phases, and the evolutionary states of massive star cores. If we assume a Salpeter IMF and $v \, / \, v_{\rm crit} \sim 0.5$, similar to our calculation in \autoref{param}, we find that a solar metallicity stellar population with a mass of $10^5$ $\mathrm{M}_\odot$ should on average contain one star that has a WNE phase while still core H burning, and that for sub-solar metallicity the mass required drops to $\approx 10^4$ $\mathrm{M}_\odot$.
\begin{figure}
\centerline{
\epsfxsize=0.5\textwidth
\epsfbox{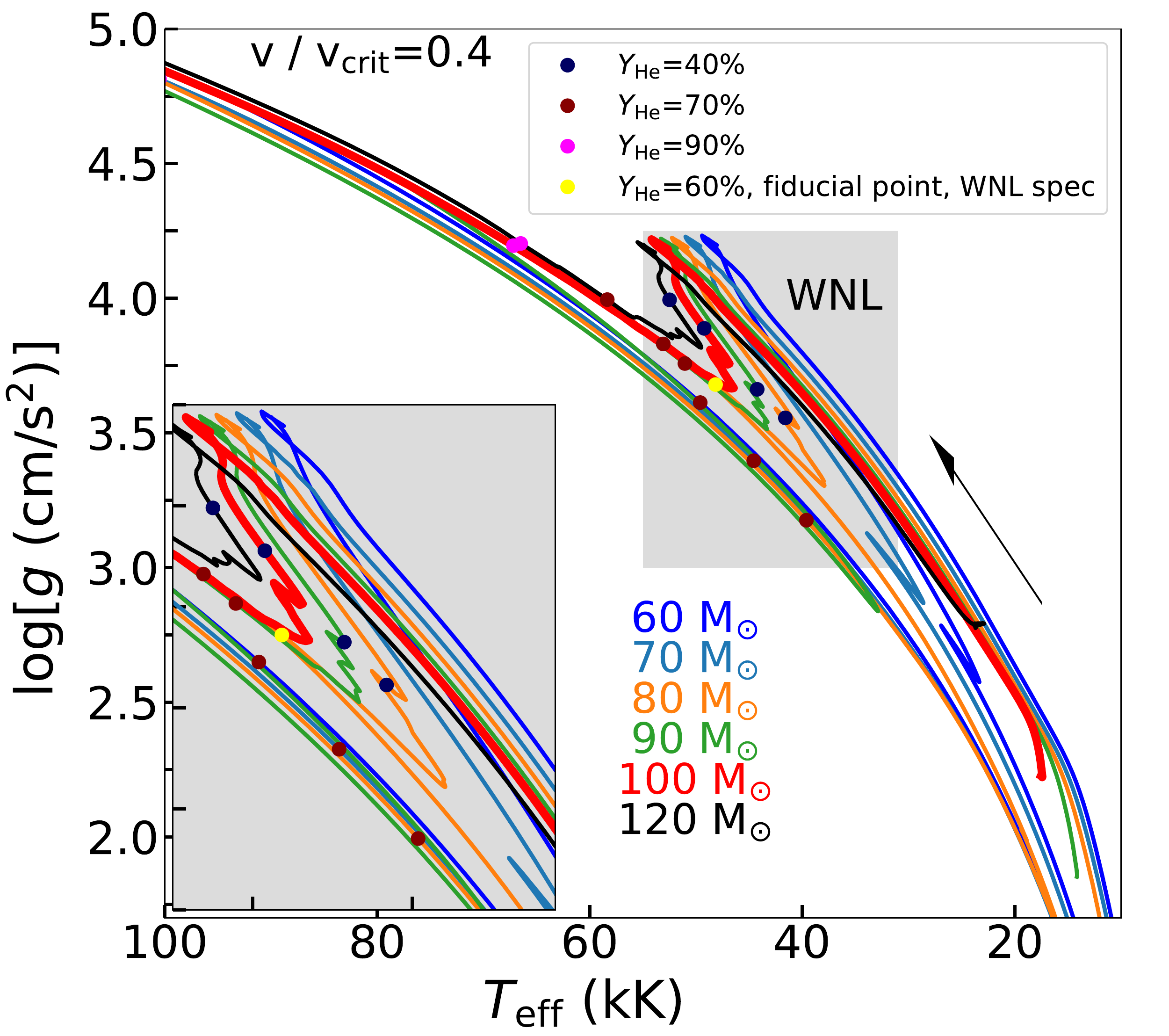}
}
\caption{Evolutionary tracks for 60, 70, 80, 90, 100 and 120 $\mathrm{M}_\odot$, as indicated, for solar metallicity and a rotation rate of $v \, / \, v_{\rm crit}=0.4$.  Dark blue, brown and magenta points mark He surface mass fractions of 40\%, 70\%, and 90\%, respectively. The grey region shows the $\log g$ $-$ $T_{\rm{eff}}$ range that characterises observed WNL stars; we also show this region zoomed-in in the inset. The yellow point shows the point in ($\log g$, T$_{\rm eff}$) space for which we calculate our fiducial spectrum (see \autoref{spectra_sec}). The arrow shows the direction of time evolution. 
}

\label{fig:g_Teff}
\end{figure}

\subsection{Direct spectral comparison}
\label{spectra_sec}

Thus far we have focused solely on surface mass-fractions as a determinant of whether a star is a WNL star. However, the WNL classification is ultimately based on spectroscopy, and stellar spectra are sensitive to surface gravity and temperature as well as abundances. In this section we verify that the stars we have identified based on abundance also have surface gravities, temperatures, and spectra consistent with those of observed WNL stars. Our spectral comparison follows the path laid out by a number of previous studies using similar methods \citep[e.g.,][]{schaerer1996, schaerer1997, grafener2002, hamann2004, hamann2006, groh2014, sander2014, ramachandran2019}.

As a first step in this direction, we show stellar tracks in $\log g$ $-$ T$_{\rm eff}$ space for 60, 70, 80, 90, 100, and 120 $\mathrm{M}_\odot$ stars with solar metallicities and  $v \, / \, v_{\rm crit}=0.4$ in \autoref{fig:g_Teff}. The navy-blue, brown and magenta points on the tracks represent the the points of 40\%, 70\% and 90\% He abundance, respectively.  The grey region shows the $\log g$ $-$ T$_{\rm eff}$ range representative of WNL stars \citep{langer2014}. We find that almost all stars with the range of surface mass-fractions (40\% and 70\%) that we have used to characterise the WNL phase also fall in the $\log g$ $-$ T$_{\rm eff}$ range that is observed for WNL stars. All the points with 90\% surface He abundances, however, representative of WNE characteristics, fall outside the grey shaded region. We therefore conclude that we not only obtain the desired surface He and N abundances, but also surface gravities and effective temperatures that match those observed in WNL stars. 

\begin{figure*}
\includegraphics[width=\textwidth]{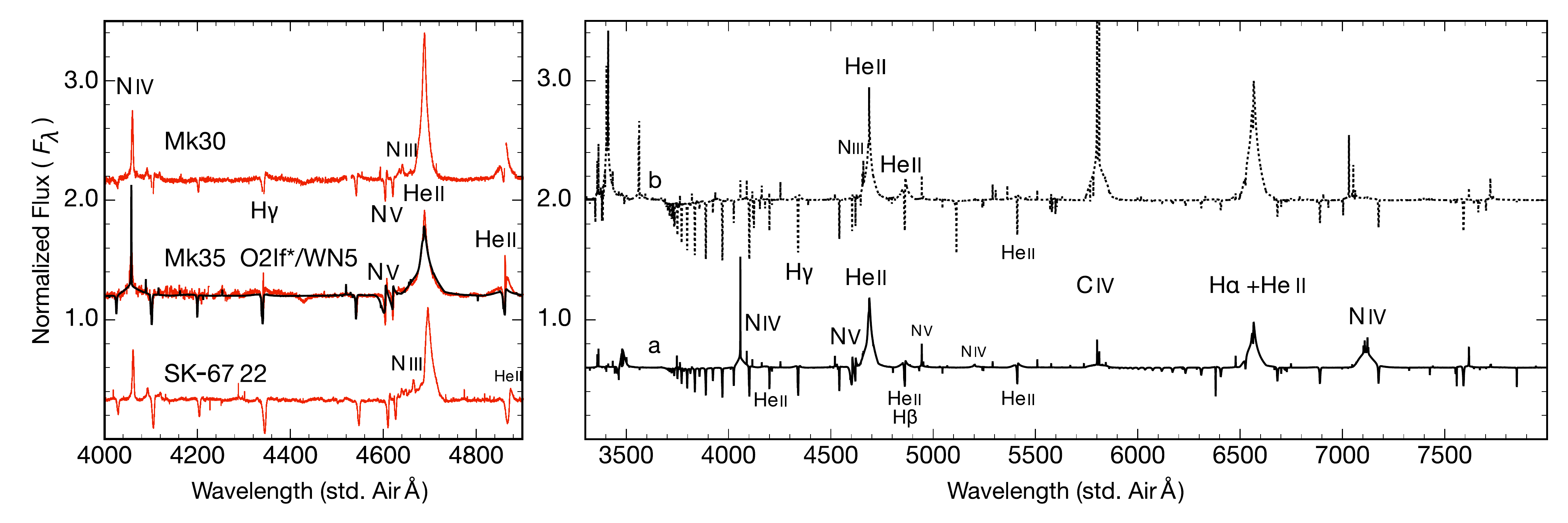}
\caption{
\label{fig:spectra}
{\bf Left:} Comparison of the spectrum predicted by CMFGEN for our $M_{\rm init} = 100$ $\mathrm{M}_\odot$ LMC metallicity star evolved with MESA to the point where the surface He abundance is $Y=0.6$, (solid black line) with the O2~If*/WNL star Melnick~35 (middle red line), also known as VFTS~545 \citep{melnick1985,evans2011,crowther2011}). No detailed fitting was attempted.  The other two red lines are MK~30 (VFTS 542) and Sk\,-67\,22 which have been usually identified as members of this class in references since the original classification of \citet{crowther2011}.
  {\bf Right:}  a) lower solid black line, a normalised $\lambda - F_\lambda$ spectrum of our model star from 3300\AA\ to 8000\AA\, showing both broad emission and narrow absorption of hydrogen, helium and nitrogen and, with relatively weak carbon lines.  b) The upper dotted spectrum is the spectrum obtained using the same stellar properties ($T_{\rm eff}$, $\log g$ and $\dot M$) as our model star, but with an un-evolved LMC surface composition.
}
\end{figure*}

To perform a direct comparison with observed stellar spectra, we make use of the VLT-Flames Tarantula Survey (VFTS) of the 30 Doradus region of the LMC \citep{evans2011, besten2014}, which provides the largest available sample of optical spectra for stars classified as WNL or a mix of WNL and another sub-type.\footnote{However, there are a number of WR stars in 30 Doradus that are not included in VFTS (Table B2 of \citet{doran2013}), notably the core WN5h stars in R136.} To facilitate the comparison we calculate the evolution of star with an initial mass of $100$ $\mathrm{M}_\odot$, $v \, / \, v_{\rm crit} = 0.4$, and $[\mathrm{Fe}/\mathrm{H}] = -0.3$, chosen to match the metallicity of the LMC; this model is not part of our overall grid, which does not have a point at $[\mathrm{Fe}/\mathrm{H}] = -0.3$, but we run this case using exactly the same setup as the models in our grid. We evolve the model until dredge-up produces a surface He fraction $Y=0.6$. This occurs after 3.04 Myr of evolution, and at this time the star remains hydrogen core burning, with a core hydrogen mass fraction of 6\%.  This places the star formally on the main sequence, although the star is in the process of developing strong mass loss and WR like features due to the (notionally unexpected for normal O-Stars) enrichment of the surface. At the point at which we calculate the spectrum, our model has the properties $M = 69.25$ $\mathrm{M}_\odot$, $L = 1.92 \times 10^6$ $\mathrm{L}_\odot$, $R_{\rm eff} = 19.905$ $\mathrm{R}_\odot$, $T_{\rm eff} = 48165$ K, $\log g = 3.68$, $v_{\rm wind} = 2995$ km s$^{-1}$, and $\dot{M} = 4.104\times 10^{-5}$ $\mathrm{M}_\odot$ yr$^{-1}$. We show the location of this star in the $(\log g, T_{\rm eff})$ plane as the yellow point in \autoref{fig:g_Teff}; clearly the star is near the middle of our proposed WNL region.

We use the 1D spherical atmosphere code CMFGEN  \citep{hillier1998, martins2004, martins2005, martins2007, martins2008}  to compute the stellar atmosphere radiation and wind properties.  CMFGEN calculates non-LTE atmosphere models with winds and includes a robust treatment of line-blanketing using the super-level approximation. It constructs a self-consistent model of the stellar fundamental properties, luminosity, radius, temperature, and hence effective gravity, mass loss, and composition, in spherical coordinates.  The transfer in these models was computed over 15 rays with a minimum of 75 radial points, out to a a radius of approximately 120 times the initial core radius.

CMFGEN is commonly run with as few as 9 elements \citep[e.g.,][]{martins2005, besten2014}, and whereas this is adequate for even detailed optical and near UV/IR analysis, it is not sufficient to get the full opacity at photon energies $>13.6$ eV. In the present work, to allow properly for radiation pressure and the emerging heavily blanketed ionising spectrum, we use a larger 20 atom model: H, He, C, N, O, Ne, Mg, Al, Si, P, S, Ar, K, Ca, Ti, Cr, Mn, Fe, and Ni, including up to 7 stages of ionisation where atomic data is available. This produces a total of approximately 13,000 levels, both in and out of grouped super--levels. Without the inclusion of such a full set of species the overall opacity and hence radiative acceleration, the extended photosphere structure, and the escaping ionising fluxes are all inadequately modelled. The effective radii and luminosities, leading to effective temperatures of nearly 50 kK, required many stages of ionisation of each species, which drove the large number of levels required to achieve a consistent solution.\footnote{One can contrast our models with those of \citet{besten2014}, who treat similar stars using CMFGEN run with a 9 atom model (H, He, C, N, O, P, Si, S, Fe) with 2724 levels. \citet{besten2014} followed a transformed mass-loss rate approach where they kept their luminosity and terminal wind speed fixed to values of $1.0\times 10^{6}$ $\mathrm{L}_\odot$ and $v_\infty = 2,800$ km s$^{-1}$ respectively. This approach allows them to scale $L$ and $v_\infty$ to other values just by tweaking the volume-filling factor ($f_{\rm v}$) of clumped winds and mass loss rates so that the transformed mass loss rate is constant. In order to cover a large range of parameter space in $L$, $f_{\rm v}$, mass loss rate and $v_\infty$, they lacked the freedom of incorporating an accurate CNO-processed abundances directly coming from stellar evolution models. In contrast, our models use surface mass-fractions taken directly from the MESA run. }

The left panel of \autoref{fig:spectra} shows our calculated CMFGEN spectrum (black line) over the spectral range $4000-4900$\AA. The emergent spectrum shares many qualities of a weak line WN5h star \citealt{hainich2014}, and the luminosity of $\approx 2\times 10^6$ L$_\odot$ is comparable to that of the more luminous WN5h stars found the centre of R136 \citep{hainich2014}. However our star has an effective temperature near 50~kK, suggesting it would be identified as a member of the relatively rare and extreme O2~If*/WNL class of so-called `slash stars' \citep{crowther2011}. Indeed, when we overlay the normalised model spectrum between 4000\AA\ and 4900\AA\ on the observed spectra of the O2~If*/WNL stars in the sample of \citet{crowther2011} (red lines in the left panel of \autoref{fig:spectra}), our model is in remarkable agreement with archetype of this stellar class: Melnick~35 (\citealt{melnick1985}; this star is referred to as VFTS~545 in \citealt{evans2011} and \citealt{besten2014}). The model spectrum is also a close match to those of two other stars in this class, Mk~30 (VFTS~542) and Sk\,-67\,22 \citep{sanduleak1970}. Defining characteristics of this class \citep{crowther2011} that are also found in our model include the strong \rion{N}{4}, moderate \rion{N}{5} lines and weak or nearly absent \rion{N}{3} emission, along with the broad \rion{He}{2} emission and narrow \rion{He}{2} and \rion{H}{2} absorption lines.

One further important point is that one obtains this excellent agreement only by calculating a model atmosphere that properly accounts for the changes in chemical composition caused by rotational mixing and exposure of the fossil convective core on the main sequence. In the right panel of \autoref{fig:spectra} we show a comparison between a CMFGEN calculation for our fiducial model and that for an otherwise identical star in which we set the atmospheric composition equal to that the star had on ZAMS, i.e., what one would obtain by neglecting the processes that cause the atmospheric composition to change over time. The comparison reveals that the change in composition dramatically affects the spectrum: only the evolved composition leads to the absence of \rion{N}{3} emission and the presence of strong \rion{N}{4} emission, two major distinguishing features of the O2~If*/WNL class.

The evolved composition also has much weaker \rion{C}{4} emission near 5800\AA. This feature is interesting to compare to the models and observations of VFTS 545 published by \citet[their Figure E38]{besten2014}. The observed star has moderate \rion{C}{4} emission, but their CMFGEN model for it predicts no \rion{C}{4} emission at all. We can now identify this discrepancy as due to the assumption of \citeauthor{besten2014} that this star is a ``classical'' WR star with an exposed core, the composition of which is fully CNO-processed and thus contains almost no carbon. In contrast, in our MESA model the surface is significantly contaminated with CNO products, but still retains a significant amount of carbon, placing it somewhere between assumptions of \citeauthor{besten2014} of full CNO processing and the assumption of a fully unprocessed surface composition represented by the dashed line in \autoref{fig:spectra}. This gives our fiducial model a moderate amount of \rion{C}{4} emission, consistent with what is observed. This comparison, together with the overall excellent spectral match between our models and the observations, allows us to identify the O2 `slash stars' as high initial mass ($>90$ $\mathrm{M}_\odot$), main sequence (albeit nearing transition), O stars whose surface compositions have been significantly modified by helium and nitrogen dredge-up, leading to the development of WR-like features. However, we do caution that this identification is subject to uncertainties in our mass loss prescription. If we adopted a prescription that increased mass loss for near-Eddington systems, it is possible that the resulting spectrum for this star would be more similar to that of a bonafide WN5h star rather than a transition O2~If*/WN5 (c.f.~Figure 11 of \citealt{crowther2010}).

\section{Discussion and implications}

\label{discussions_sec}

Thus far we have discussed different physical mechanisms to produce WNL stars and their spectra. In this section, we discuss the broader implications of our findings, as well as adding some caveats.

\subsection{Spectra and ionising photon budget in high-redshift star clusters}

In the high redshift universe, stars are mostly metal poor, and we have shown in  \autoref{results_sec} that such stars will spend the majority of their lives with WNL-like rather than O-like characteristics if they have moderate ($v \, / \, v_{\rm crit} \sim 0.25$) or faster rotations rates. Theoretical models suggest that most massive stars will be rotating at least this fast at birth, independent of metallicity \citep{Rosen12a}. This suggests that the ionising spectra of high-redshift, low-metallicity galaxies should be more WNL-like than O-like.

To our knowledge, no spectral synthesis codes in wide use in the astronomical community properly account for this shift. While a number of authors have considered the effects of rotation and binarity on spectral synthesis \citep[e.g.,][]{levesque2012, eldridge2017}, even these recent calculations implicitly assume that stars retain their original surface composition until they leave the main sequence, yielding spectra that resemble Case (b) rather than Case (a) in \autoref{fig:spectra}. These spectra differ very little in their hydrogen-ionising photon fluxes, and thus the effect of this assumption on the total ionising photon budget is minimal. However, there are much larger changes at higher energy. The spectrum we compute for our model star with a self-consistently evolved surface composition (Model a in \autoref{fig:spectra}) contains a factor of $\approx 20$ fewer photons at the \rion{He}{2} and \rion{O}{3} edges than the ``control'' case of a star with identical bulk properties but an unevolved composition (Model b in the figure). Determining the full impact of this change, and its importance for spectral line diagnostics that depend on these species, will have to await a full spectral synthesis calculation, however, since the stars we are modelling here are may be sub-dominant compared to ``classical'' WR stars or those produced by envelope stripping in tight binaries.

Future cluster spectral synthesis codes will need to take the helium enrichment of the high end of the HR diagram into account in order to get accurate composite spectra at energies beyond a few Rydberg. The extreme `slash star' phase \citep{crowther2011} shown in our spectral model in \autoref{fig:spectra}, characterised by very powerful winds when the star is still on the main sequence, is relatively short. Our modelling shows however that \textit{all} the main sequence stars above $\sim 90$ $\mathrm{M}_\odot$ will be spend a significant portion of their lives with surfaces that are He and N enriched. They will, therefore, have different surface opacity-age relationships and will contribute a different radiation spectrum than would be expected if the surface composition were un-evolved, as assumed in current spectral synthesis codes. Full quantitative evaluation of the impact of this change will require a large library of atmospheres covering the full range of WNL physical properties and atmospheric compositions, which we leave to future work. We also remind the reader again that the precise range in time and initial stellar mass at which surface He and N enrichment occur is subject to uncertainties in stellar mass loss prescriptions.

\subsection{N-enrichment in the early universe}

WNL stars have high surface N mass-fractions, or rather high [N/Fe] or [N/C], as well as high He. The ubiquity of such stars at low metallicity has potentially-important implications for the long-standing problem of the abundance of N in high-redshift galaxies.
 \citet{dopita2016} show that at low oxygen abundances [O/H], the N/O ratio 
has very large scatter but shows a clear plateau, rather than increasing linearly with [O/H] as is found at higher [O/H] values. This is strongly indicative of a source of primary N. Further evidence for enhanced N at early times comes from the surveys of     
\citet{masters2014} and \citet{shapley2015}, who observed a sample of $z\sim 2$ galaxies and found that they have elevated [N~\textsc{ii}]/H$\alpha$ line ratios at fixed [O~\textsc{iii}]/H$\beta$ compared to local galaxies. These authors propose that the offset can be explained only if there are an unusually high number of classical WR stars ejecting their N-rich envelopes, which are subsequently ionized, yielding elevated [NII] emission. This scenario, however, runs into two significant problems. First, most stellar evolution models suggest that classical WR stars should be rare or non-existent at low metallicity, because metal-poor stars do not lose enough mass to reveal their N-rich cores. Second, the timing of this scenario is difficult to arrange. Classical WR stars do not appear until late in the evolution of a stellar population, when the ionizing luminosity has already dropped significantly. Thus, a scenario in which classical WR stars are responsible for enhanced [N~\textsc{ii}] lines is possible only if the ionizing photons are coming not from the stars that are producing the N, but from a subsequent generation of stars that formed later, on a timescale long enough for the N to have been produced, but short enough for the extra N not to have been diluted by mixing with the ISM.

A scenario in which N pollution of high-redshift nebulae arises from WNL stars rather than classical WR stars neatly sidesteps both these problems. WNL stars are expected to be ubiquitous even at low metallicity, and, though they do not lose as much mass as classical WR stars, the mass they do lose is very rich in N. Moreover, this pollution begins in the first $\sim 1$ Myr after a stellar population forms, well before the ionizing luminosity begins to drop, removing the sequencing problem. We investigate this scenario for self-pollution of high-z nebulae by the most massive O stars in a follow-up paper \citep{roy2019}.

\subsection{Caveats}
No paper on the evolution of massive stars would be complete without a discussion of caveats. Here we highlight two: abundance inconsistencies and binarity. With regard to the former,
 there is a serious inconsistency in our atmosphere models, which are precomputed and thus are not precisely matched to the evolving composition that develops as a result of the transformation to WNL-like surface mass-fractions. This is unlikely to affect wind mass loss rates, since these are  mainly sensitive just to the amount of Fe per unit mass, which is not altered by the  evolving surface abundances \citep{vink2001}. The transformation from O-like to WNL-like surface mass fractions will, however, affect total stellar luminosities and surface temperatures, which could in turn have minor but not negligible effects on stellar evolution. Ideally one should self-consistently evolve the abundances in all three places they matter: atmospheres, interior opacities, and nuclear reaction rates.


Our second caveat is that, at least in the local Universe, most massive stars are members of binary systems that are close enough to interact at some point during stellar evolution \citep{sana2012}. In these cases, mass transfer via Roche lobe overflow or common envelope (CE) evolution may  provide an additional mechanism for stripping off the outer layers of massive stars that are too metal-poor to lose much mass on their own. Therefore binarity may enhance the  early onset of high surface mass-fractions of He and N compared to what we have found in our single-star models. To address this possibility, one must study binary evolution along with single star evolution \citep[e.g.,][]{eldridge2009, eldridge2017, stanway2016, xiao2018, gotberg2019}. Current models suggest that binaries dominate the overall ionising flux, but that they do not begin to do so until stellar populations reach ages $\gtrsim 5$ Myr. Thus at the young ages when the effects we study in this paper are most prominent, binaries are expected to be sub-dominant. However, we caution that the results of any such study are likely to be sensitive to parameter choices such as the initial distribution of orbital periods and mass ratios, which are almost completely unconstrained by observations in the high-redshift, low-metallicity environments where we have argued that WNL stars are likely to have their greatest impacts.

\section{Conclusions}
\label{conclusions_sec}
In this paper, we study the detailed structure of massive stars ($M\geq 60$ $\mathrm{M}_\odot$) across a wide range of metallicity, mass, and rotation rate, with an emphasis on the various  physical mechanisms that transport heavy elements from the core to the surface.

The primary findings of the paper are:
\begin{itemize}
\item Massive stars evolve to surface mass-fractions that are a good match to the WNL class, defined by considerable surface He ($\geq$ 40\%) and N ($\sim 10- 30\times$-initial N abundance) enhancements on the main sequence, at all metallicities if they are rotating at $\gtrsim 30\%$ of breakup, and at near-solar metallicity even if they are non-rotating.
\item  He and N surface mass-fractions rise in parallel, driven by three primary physical mechanisms. (i) During the first $\approx 1-2$ Myr of evolution, rotation-induced mixing raises the surface He abundance to $40- 50$\% in massive ($M\gtrsim 100$ $\mathrm{M}_\odot$), moderately to highly rotating ($v \, / \, v_{\rm crit}\gtrsim 0.3$) stars. (ii) After $\sim 1$ Myr, moderately metal-rich massive stars lose enough mass via main sequence winds to begin exposing ``fossil'' convective cores -- parts of the star that are not currently in the convective core, but are He- and N-enhanced because they were part of the core when the star first joined the main sequence. This process occurs even in non-rotating stars, and for a 100 $\mathrm{M}_\odot$, solar-metallicity star will raise the surface He abundances to $\approx 60\%$ by $\approx 2.5$ Myr of evolution.
(iii) Towards the end of and after the core H depletion, stars experience the ``classical'' Wolf-Rayet phase, during which the star approaches the Eddington limit and experiences rapid mass loss. This enhances surface He abundance to $90- 95$\%.
\item The second mechanism for enhancing surface He and N -- exposure of the fossil convective core -- provides a natural explanation for the origin of slowly-rotating WNL stars \citep{herrero2000, vink2017}, without the need to invoke exotic scenarios for stellar spin-down. However, we caution that the importance of this mechanism is subject to uncertainties in mass loss prescriptions for very massive stars.
\item Assuming a Salpeter IMF, and that massive stars are typically born rotating at $v \, / \, v_{\rm crit} \gtrsim 0.4$, a cluster of $\approx 10^4$ $\mathrm{M}_\odot$ (for solar metallicity) or $\approx 10^3$ $\mathrm{M}_\odot$ (for sub-solar metallicity) is large enough to on average produce one star that will experience a WNL phase at some point during its life.

\item Contrary to the hypotheses advanced by some previous authors \citep[e.g.,][]{schaerer1992, crowther2007}, there is no one-to-one connection between WN-subtypes (WNE and WNL) as defined by surface mass-fractions and the evolutionary phases of stellar cores. Depending on the stellar mass, metallicity, and rotation rate, both phases straddle the core H depletion phase; for some stars, a substantial fraction of the time when they show WNE-like surface mass-fractions coincides with core H burning, and for others a good fraction of the time they spend with WNL-like abundances is after core H burning has ceased.   
\item We show that the stars we identify as candidate WNLs based on their surface mass-fractions also match the observed surface gravity and temperatures for WNL stars, and that their spectra also agree well with observations of WNL star spectra. This gives us confidence that we have indeed identified the evolutionary pathways that lead to WNLs.
\item The evolution of surface composition associated with the transition from O to WNL has relatively little effect on the total hydrogen-ionising photon budget, but causes very significant changes in the ionising photon flux beyond a few Rydberg. The effects on He~\textsc{ii} and O~\textsc{iii} ionising fluxes are particularly strong, with model atmospheres calculated for the evolved surface composition of WNL stars producing a factor of $>20$ less emission than comparable models that neglect evolution of surface mass-fractions. This effect is not included in most spectral synthesis codes, which generally assume no evolution of surface mass-fractions until stars depart the main sequence. This effect may necessitate the re-evaluation of some spectral line diagnostics, particularly those based on oxygen lines.
\end{itemize}

\section*{Acknowlegements}
Authors acknowledge the valuable and constructive comments from the anonymous referee. AR and MRK acknowledge support from the Australian Research Council's \textit{Discovery Projects} and \textit{Future Fellowship} funding scheme, awards DP160100695 and FT180100375. RSS and MAD acknowledge the support of the Australian Research Council Centre of Excellence for All Sky Astrophysics in 3 Dimensions (ASTRO 3D), through project number CE170100013. AR acknowledges Charlie Conroy, Aaron Dotter and Jieun Choi for providing the MIST-II setup and many useful discussions. AR gratefully acknowledges the support of Lisa Kewley's ARC Laureate Fellowship (FL150100113). AH has been supported, in part, by a grant from Science and Technology Commission of Shanghai Municipality (Grants No.16DZ2260200) and National Natural Science Foundation of China (Grants No.11655002), and by the Australian Research Council through a Future Fellowship (FT120100363). This work benefited from support by the National Science Foundation under Grant No. PHY-1430152 (JINA Center for the Evolution of the Elements). This research/project was undertaken with the assistance of resources and services from the National Computational Infrastructure (NCI), which is supported by the Australian Government. We acknowledge constructive comments from Jorick Vink, Joachim Bestenlehenr,  JJ Elridge and Wolf-Rainer Hamann.

\appendix
\section{Parameter sensitivity}
\label{param_sensitivity}

In this Appendix we explore the sensitivity of our results to uncertain parameters describing the rates of chemical and angular momentum transport in stars.

\subsection{Rotational instabilities}
\label{conv_rot}
\begin{figure*}
\centerline{
\epsfxsize=1.0\textwidth
\epsfbox{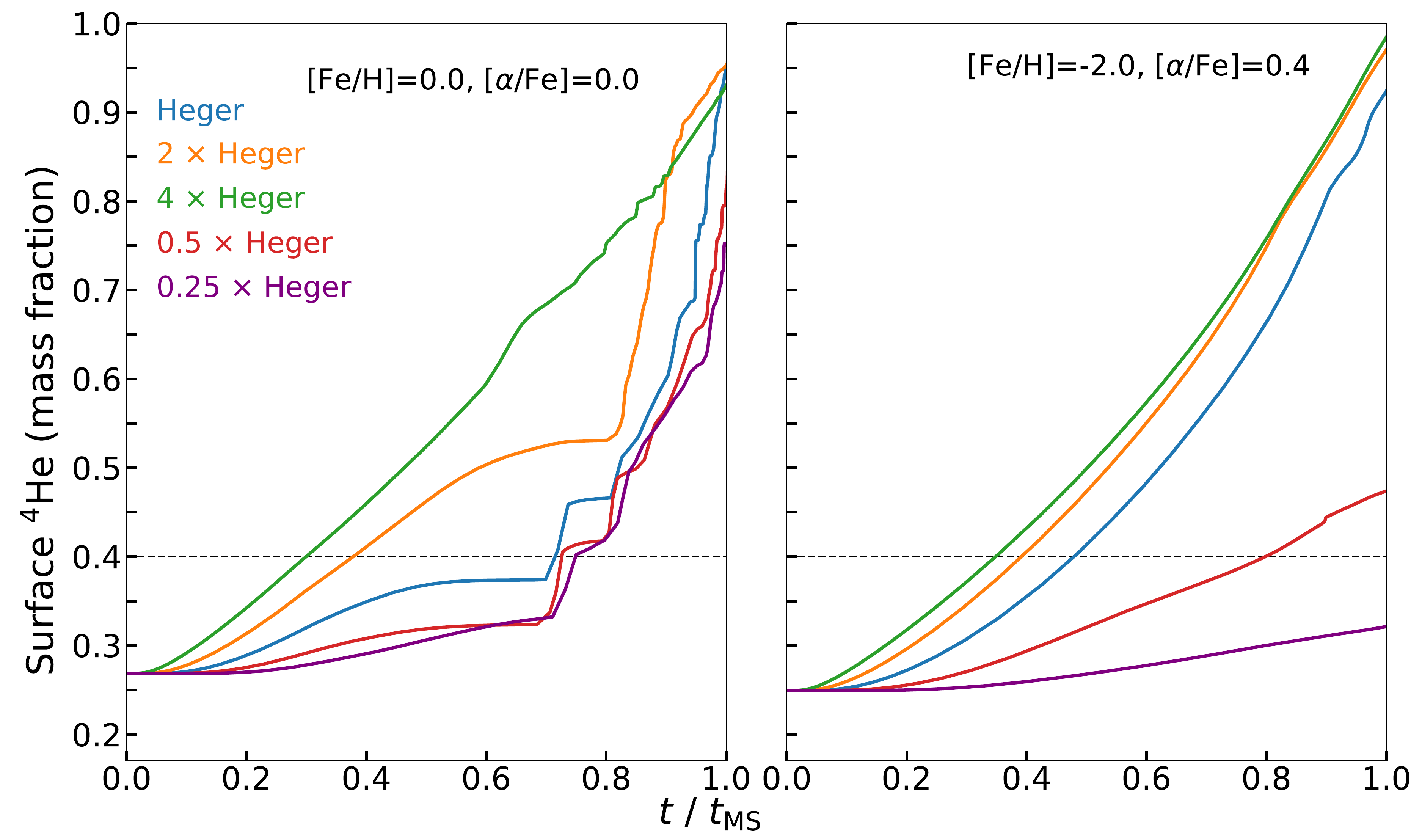}
}
\caption{Time evolution of surface helium abundances for varying chemical and angular momentum diffusion coefficients at two metallicities, [Fe/H]=0.0 (left panel) and -2.0 (right panel), for 100 $\mathrm{M}_\odot$ and for [$v \, / \, v_{\rm crit}$]=0.4. The black-dashed horizontal line shows the surface helium mass fraction of 40\% that we use to roughly delineate the transition from O to WNL. In this figure, the blue line labelled ``Heger'' in the legend corresponds to adopting the diffusion coefficients $D$ and $\gamma$ recommended by \citet{heger2005}, as we do in the main text. Other colours show results of increasing or decreasing these coefficients by the amounts stated in the legend.
}
\label{fig:heger}
\end{figure*}
We use the models of \citet{heger2005} for the chemical and angular momentum diffusion coefficients due to rotational mechanisms, $D$ and $\gamma$, in \autoref{results_sec}.  
In \autoref{fig:heger}, we show the time evolution of the He surface abundance for two cases (mass 100 $\mathrm{M}_\odot$, $v \, / \, v_{\rm crit} = 0.4$, [Fe/H]$=0$ and $-2$) where we have increased or decreased these coefficients by factors of 2 and 4 relative to the value recommended by \citet{heger2005}, but leaving all other parameters unchanged. As the figure shows, for the metal-rich case ([Fe/H]$=0$) changing the diffusion coefficients within this range leads to tens of percent differences in the relative timings and enrichment levels achieved in phases `AB' and `BC' (see \autoref{mix}). This translates to tens of percent differences in the durations of the WNL and O phases; however, there is a lower limit on the amount by which reducing the rotational mixing coefficient can delay on onset of a WNL phase, since eventually mass removal will force the surface He abundance upward. The metal-poor case ([Fe/H]$=-2$) is more sensitive to the choice of rotational mixing parameters, since mass loss via winds is much less important for it. In this case, a sufficiently extreme choice for the rotational mixing coefficient ($1/4$ of the fiducial value) can prevent a WNL phase from happening at all. We therefore conclude that duration of the WNL phase is at least moderately sensitive to the exact value of the rotational mixing coefficient, with the sensitivity being greater at low metallicity.

\subsection{Overshoot mixing}
\label{conv_ovr}
\begin{figure}
\centerline{
\epsfxsize=0.5\textwidth
\epsfbox{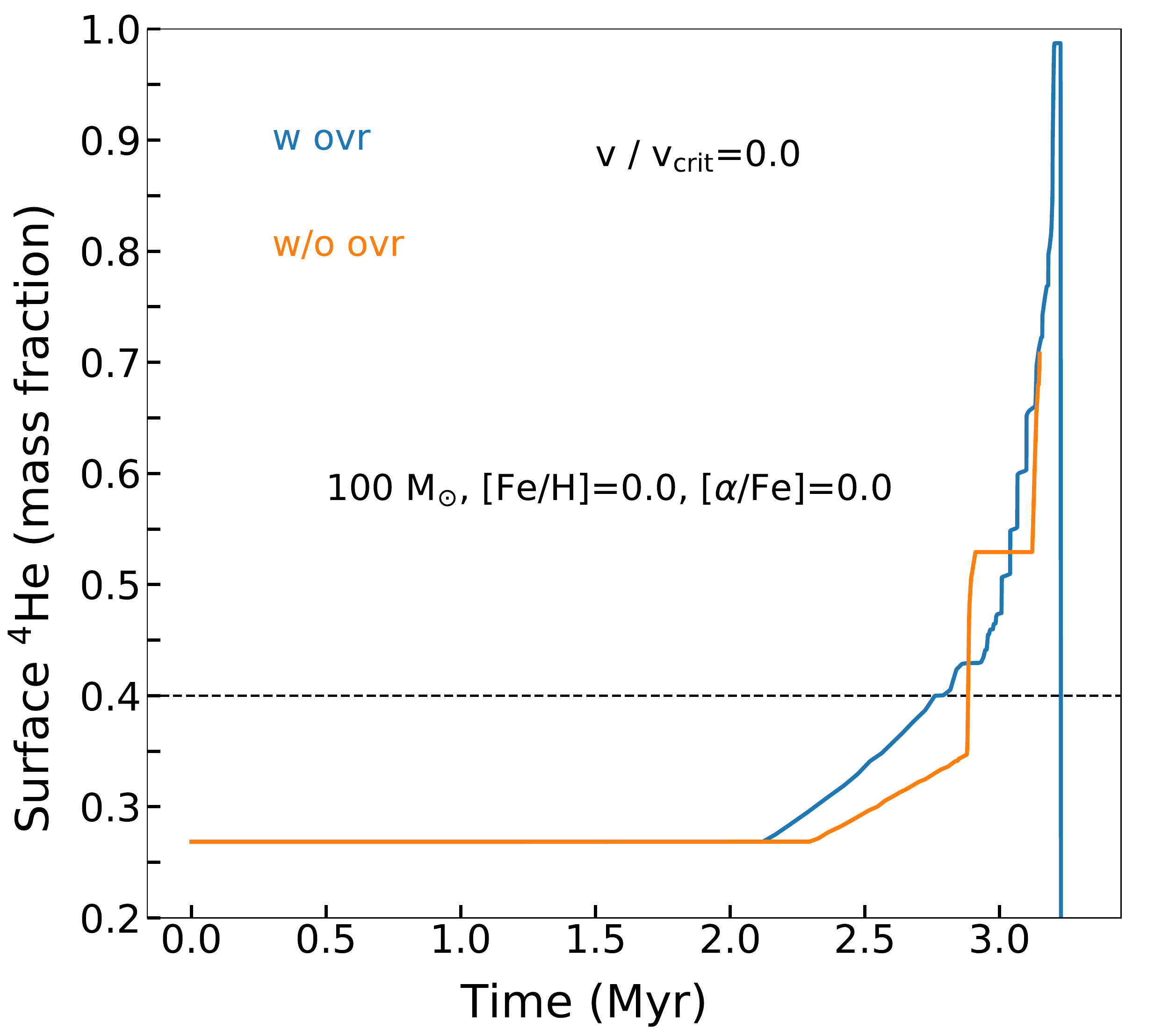}
}
\caption{Time evolution of the surface helium abundance for $v \, / \, v_{\rm crit}$=0.0 and for a 100 $\mathrm{M}_\odot$, solar metallicity star with and without any overshoot mixing. The black dashed line shows the He fraction of 40\% we use to roughly delineate the transition from O to WNL.
}
\label{fig:ovr_conv}
\end{figure}
Overshoot-convective mixing is one of the non-rotational mixing mechanisms as described in \autoref{mixing_subsec}.  Our fiducial treatment of overshoot mixing follows \citet{choi2016}, but the mixing rate is substantially uncertain. To test the dependence of our results on this choice, in \autoref{fig:ovr_conv} we show the time evolution of the surface He abundance of non-rotating 100 $\mathrm{M}_\odot$ star with solar metallicity with and without the inclusion of overshoot mixing. We focus on this case because it is the one for which overshoot mixing is mostly likely to be important, since there is no rotational mixing. 
The figure shows that the inclusion of overshoot mixing has relatively modest effects on the results. It changes the exact timing for the onset of phase `BC', and the exact shape of the surface He increase during it, but does not change the qualitative result, and makes a quantitative difference in the He surface abundance of only $\sim 20\%$.

\subsection{Mass-loss prescriptions}
\label{mass_loss_subsec}
\begin{figure*}
\centerline{
\epsfxsize=1.0\textwidth
\epsfbox{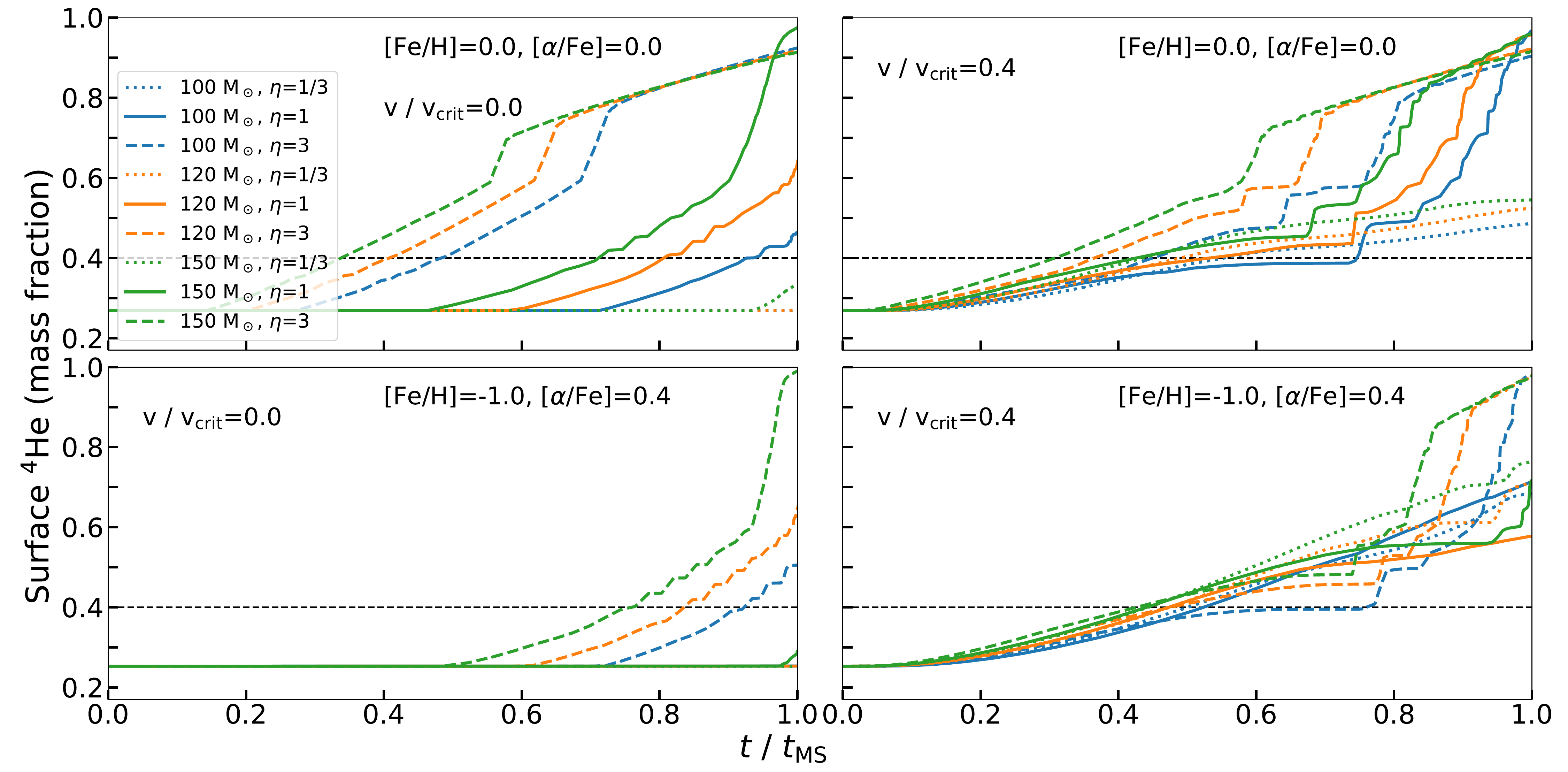}
}
\caption{Time evolution of surface He mass-fraction during the MS for different mass-loss prescriptions, $\eta=$ 1/3 (the dotted lines), 1.0 (the solid lines) and 3.0 (the dashed lines), for three masses, 100 (the blue lines), 120 (the orange lines), 150 $\mathrm{M}_\odot$ (the green lines), for two metallicities, [Fe/H]=0.0 (the top panels) and -1.0 (the bottom panels), 
 and for two rotation rates, $v \, / \, v_{\rm crit}$=0.0 (the left panels) and 0.4 (the right panels). The black dashed lines show a surface He mass-fraction of 40\%. 
}
\label{fig:massloss}
\end{figure*}

Our fiducial prescription for mass loss is as described in \autoref{ssec:winds}. Here we explore the sensitive of our results to this choice by repeating twelve test cases -- [Fe/H]$=0$ and $-1$, $v \, / \, v_{\rm crit} = 0$ and $0.4$, and $M = 100$, 120, and 150 $\mathrm{M}_\odot$ -- with mass loss rates multiplied by a factor $\eta$.
We show the time evolution of surface He mass-fraction from these tests in \autoref{fig:massloss}.

The main conclusion to be drawn from \autoref{fig:massloss} is that, as one might expect, varying the mass loss rate has very strong effects for non-rotating stars. Tripling the mass loss rate allows surface He enrichment and a WNL phase to occur even for [Fe/H]$=-1$, while reducing it by a factor of 3 eliminates the WNL phenomenon entirely even at solar metallicity. The effects for rapidly rotating stars are less dramatic, since rotational mixing provides a mechanism to bring He to the surface that is independent of mass loss. The main effect of increasing or decreasing the mass loss rate in this case is to change the maximum level of He enrichment, with cases of reduced mass loss never exceeding $\sim 50-60\%$ He on the main sequence, while enhanced mass loss cases reach $\sim 90\%$ surface He.

\section{Convergence tests}
\label{conv_test}
\begin{figure*}
\centerline{
\epsfxsize=1.0\textwidth
\epsfbox{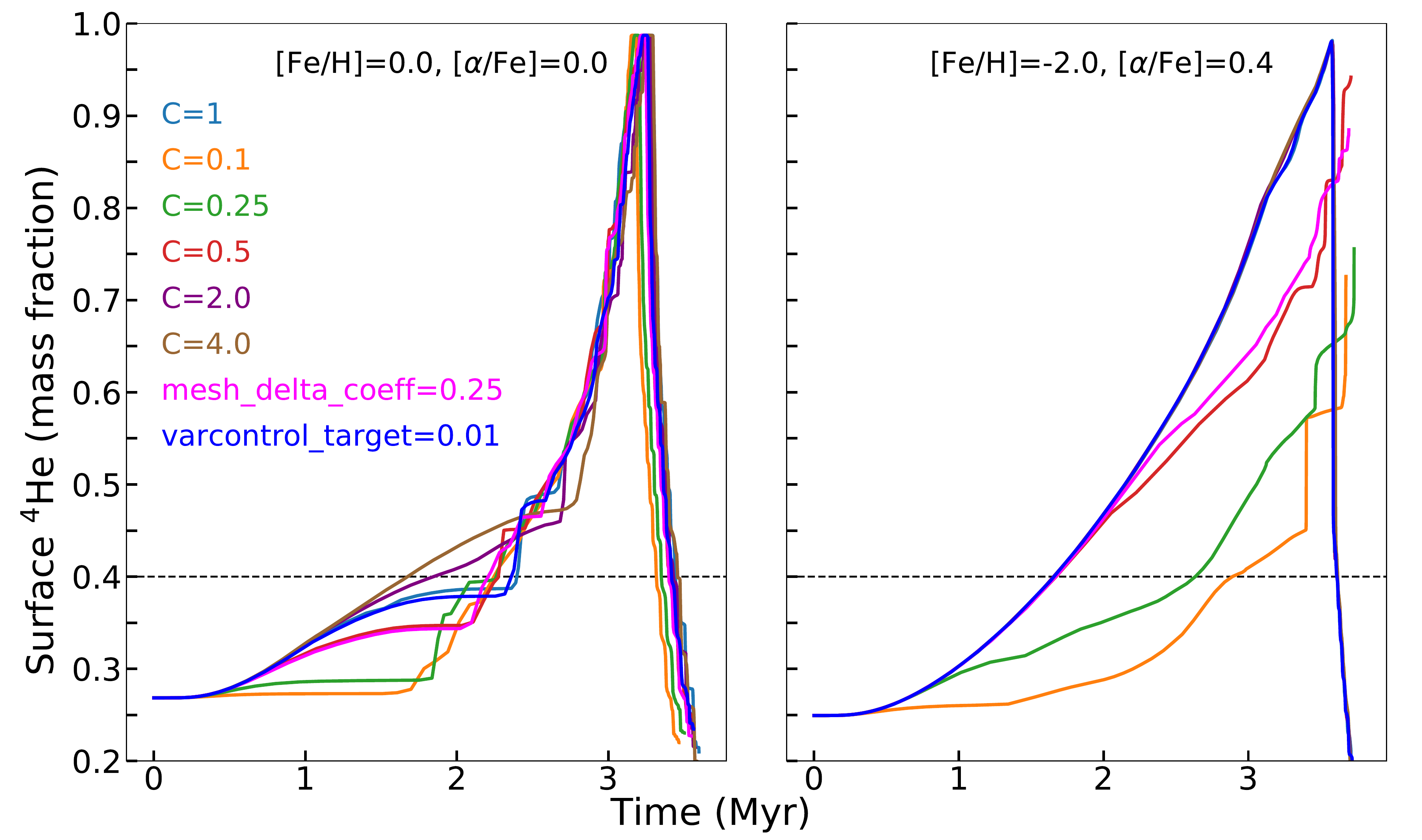}
}
\caption{The time evolution of surface He mass-fraction for a sample star of 100 $\mathrm{M}_\odot$, $v \, / \, v_{\rm crit}=0.4$ and solar metallicity, run with different numerical tolerance parameters as indicated in the legend. C is a constant factor that multiplies the default values of both the spatial (\texttt{mesh\textunderscore delta\textunderscore coeff}$=0.5$) and temporal (\texttt{varcontrol\textunderscore target}$=1\times 10^{-4}$) resolution. Line-styles and line-colours are described in the legend of the figure.
}
\label{fig:res_convergence}
\end{figure*}

\begin{figure*}
\centerline{
\includegraphics[width=1.0\textwidth]{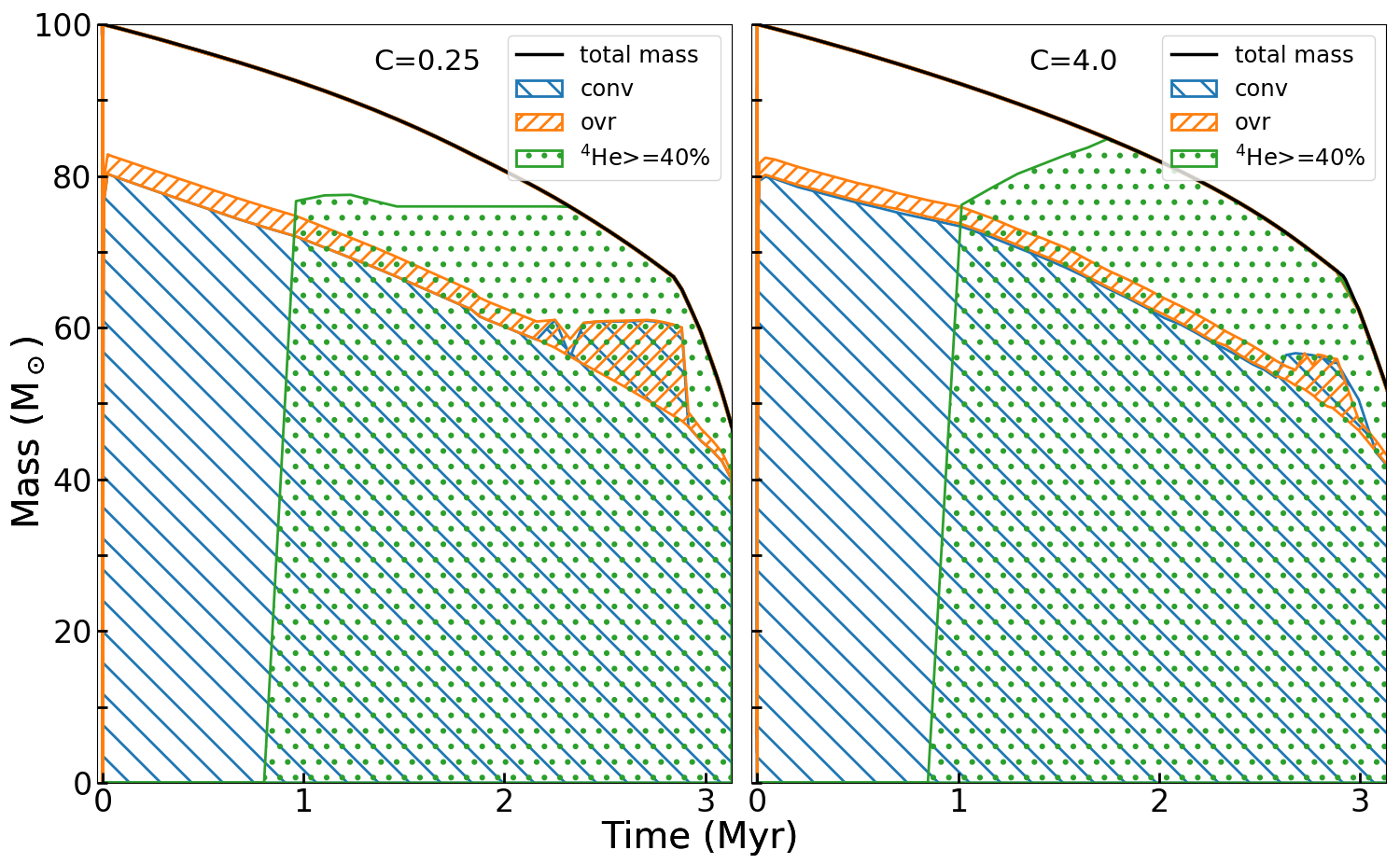}
}
\caption{Kippenhahn diagram of a sample star, 100 $\mathrm{M}_\odot$ star with $v \, / \, v_{\rm crit}=0.4$ at solar metallicity, for two values of numerical resolution, \texttt{C} = 0.25 (the left panel) and 4.0 (the right panel).
}
\label{fig:kipp_res_comp}
\end{figure*}

As a MESA simulation runs, it adapts the spatial and temporal resolution based on tolerances for the changes in variable values across mesh cells and over time. Spatial resolution is controlled by the parameter \texttt{mesh\textunderscore delta\textunderscore coeff}, which specifies the maximum fractional change in quantities between adjacent cells; temporal resolution is controlled by \texttt{varcontrol\textunderscore target}, which specifies the maximum average fractional change in variables per time step. Both of these parameters can be adjusted independently, or multiplied by an overall control parameter \texttt{C}. See \citet{paxton2011} for a full description of these parameters and how they are implemented. The default values we use in the main text, which match those used in \citet{choi2016}, are \texttt{mesh\textunderscore delta\textunderscore coeff}$=0.5$, \texttt{varcontrol\textunderscore target}$=10^{-4}$, and \texttt{C}$=1$.

To test the sensitivity of our results to these parameters, we re-ran the cases $M=100$ $\mathrm{M}_\odot$, $v \, / \, v_{\rm crit} = 0.4$, and [Fe/H]$=0$ and $-2$ with values of \texttt{C} from 0.1 to 4.0; we also independently varied the spatial and temporal resolution parameters, using \texttt{mesh\textunderscore delta\textunderscore coeff}$=0.25$ and \texttt{varcontrol\textunderscore target}$=0.01$. We show the results of these experiments in \autoref{fig:res_convergence}. The primary conclusion to be drawn from this figure is that variations in the temporal resolution parameter make little difference to the outcome, but that increasing the spatial resolution has effects that are qualitatively very similar to those of reducing the rotational mixing coefficients (\autoref{conv_rot} and \autoref{fig:heger}): both lower rotational mixing coefficients and higher spatial resolution reduce the amount of rotationally-driven transport of He to the stellar surface, and thus reduce the importance of phase `AB' in enriching stellar surfaces. The overall change in when stars evolve into WNLs is fairly modest for metal-rich stars, since for these stars exposure of the fossil convective core will take over even if rotational mixing is suppressed; for metal-poor stars that lack significant winds, the effects are much larger. The similarity between the effects of changing the spatial resolution and changing the diffusion coefficients suggests that chemical transport in our simulations can be increased substantially by numerical diffusion; confirm this intuition in \autoref{fig:kipp_res_comp}, which shows Kippenhahn diagrams produced in otherwise-identical simulations run with \texttt{C} $= 0.25$ and $4.0$, respectively. The two cases are very similar in most properties, including when regions of $>40\%$ He appear, but in the lower resolution case the He is transported upwards through the star more rapidly than in the higher resolution case.

While this result might initially appear to suggest that all our results on rotational mixing should be viewed with suspicious, since they are not converged, it is important to understand that the rotational mixing rate coefficients we are using are not themselves independently-calculated quantities whose absolute values are known as a result of a separate physical calculation. Instead, the rate coefficients themselves have been tuned by comparison to observations. Thus the correct interpretation of our result is that the tuning process is resolution-dependent, i.e., if one varies \texttt{C}, then the rotational mixing coefficient $D$ and $\gamma$ should also be varied, so as to maintain the same effective rate of rotational transport.

\bsp	
\label{lastpage}
\end{document}